\documentclass[a4paper,12pt]{article}
\usepackage[margin=2.5cm,nohead]{geometry}
\usepackage[utf8]{inputenc}
\usepackage{natbib}
\usepackage{color}
\usepackage{sgame}
\usepackage{pdflscape}
\usepackage{graphicx}
\usepackage{amssymb}
\usepackage{amsbsy}
\usepackage{amsmath,amsthm}
\usepackage{booktabs}
\usepackage{rotating}
\usepackage{colortbl}
\usepackage{caption}
\usepackage{setspace}
\usepackage{wasysym}
\usepackage{longtable}

\usepackage{subcaption}
\usepackage{float}
\usepackage{graphicx}
\usepackage{soul}
\usepackage{hyperref}
\hypersetup{
    colorlinks=true,
    linkcolor=blue,
    filecolor=blue,      
    urlcolor=blue,
    citecolor = blue,
}

\usepackage[nobottomtitles]{titlesec}
\newenvironment{myepigraph}
  {\par\hfill\itshape
   \begin{tabular}{@{}r@{\hspace{2em}}}} 
  {\end{tabular}\par\medskip}

\begin{document}
\sloppy
\title{Competing for Attention -- The Effect of Talk Radio on Elections and Political Polarization in the US
}

\author{Ashani Amarasinghe\thanks{%
University of Sydney and SoDa Laboratories, Monash University; email:
ashani.amarasinghe@monash.edu} \hspace{5mm} Paul A.~Raschky\thanks{%
Department of Economics  and SoDa Laboratories, Monash University; email: paul.raschky@monash.edu.} }
\date{June 2022}
\maketitle
 

\begin{abstract}
\medskip 
\noindent This paper studies the effects of talk radio, specifically the Rush Limbaugh Show, on electoral outcomes and attitude polarization in the U.S. We propose a novel identification strategy that considers the radio space in each county as a market where multiple stations are competing for listeners' attention. Our  measure  of  competition is a spatial Herfindahl-Hirschman Index (HHI) in radio frequencies. To address endogeneity concerns, we exploit the variation in competition based on accidental frequency overlaps in a county, conditional on the overall level of radio frequency competition. We find that counties with higher exposure to the Rush Limbaugh Show have a systematically higher vote share for Donald Trump in the 2016 and 2020 U.S. presidential elections. Combining our county-level Rush Limbaugh Show exposure measure with individual survey data reveals that self-identifying Republicans in counties with higher exposure to the Show express more conservative political views, while self-identifying Democrats in these same counties express more moderate political views. Taken together, these findings provide some of the first insights on the effects of contemporary talk radio on political outcomes, both at the aggregate and individual level.\\\medskip 


\medskip

\end{abstract}
\pagebreak

\begin{myepigraph}
 How did you start turning away from Rush Limbaugh?\\
In the late 90s, just playing around the radio dial, [\dots] \\
 I found a very very funny show,  it was \\
``Wait Wait \dots Don't Tell Me!'' on NPR.\\[1.5ex]
Interview with a former Rush Limbaugh fan.\\
The Brainwashing of My Dad (2015)  
\end{myepigraph}
\doublespacing
\section{Introduction}

\noindent The events around the 2016 and 2020 U.S. presidential elections have ignited a wide discussion about the role of populist media channels in influencing election outcomes. This discussion has been supplemented by recent academic evidence that exposure to populist media systematically increases the Republican vote share \citep[e.g.,][]{DellaVigna07,martin17,ash21}. A large portion of the existing literature in political economy and political science has its focus on the modern forms of media, such as the internet, social media and the television, in propagating populist sentiments in the U.S. \citep[e.g.,][]{Gentzkow06,Gentzkow11,campante13,melnikov21}.\footnote{For effects of television, internet and social media on populist sentiments in other countries, see, for example, \citet{durante12}, \citet{campante17}, \citet{peisakhin18} or \citet{durante19}.}

Interestingly however, \citet{boxell17} document that the recent increase in polarization is mainly observed among those older than 65 years which is the demographic group least likely to use the internet and social media\footnote{Pew Research Center, \textit{Internet/Broadband Fact Sheet 2021}, https://www.pewresearch.org/internet/fact-sheet/internet-broadband/}. This then begs the question what other forms of media exposure exacerbate the spread of populist ideologies. Political commentators have argued that a more traditional form of media, talk radio, has been the breeding ground for a more populist form of political discourse in the U.S., long before the emergence of populist TV channels, the internet and social media. While Wang (2021) studies the effects of populist radio with a \textit{historical} perspective, a systematic, empirical analysis of the impact of talk radio on \textit{contemporary} U.S. elections is absent in the literature. 

The purpose of this paper is to fill this gap by estimating the effect of talk radio on electoral outcomes in the U.S. We focus specifically on the key conservative talk radio show in the U.S., the Rush Limbaugh Show, which commenced in 1988 and aired consistently until 2021, up to the death of Rush Limbaugh, the host. We first construct a county-level measure of the exposure to the Rush Limbaugh Show (hereafter ``the Show"), based on geo-referenced data on radio frequency contours of all U.S. radio stations that aired the Show. We then combine this measure with county level election results from the U.S. Presidential Elections from 1980 to 2020, as well as individual-level survey data on attitude towards key social issues, to examine whether exposure to the Show had a systematic effect on (a) the Republican vote share and (b) attitude polarization.

The major empirical challenge however, is that the Show was mainly broadcasted via AM radio frequencies, making it available almost everywhere in the continental U.S. Therefore, it is not possible to apply standard identification strategies that either rely on exogenous spatial variation in radio signal availability  \citep[e.g.,][]{olken09}, or the staggered roll-out of a particular media program \citep[e.g.,][]{DellaVigna07}. We therefore propose a novel identification strategy based on the idea that there is competition for radio listener attention. 

We argue that the degree to which listeners within a particular county are exposed to the Show depends not only on the (endogenous) number of contours broadcasting the show, but also on the number of alternative radio programs available in the county. Accordingly, we view the radio space in each county as a market where multiple stations are competing for listeners' attention. We consider FM stations, which primarily deliver entertainment and musical programs, as the key competitor to AM stations delivering  the  Show. A larger amount of other radio options increases the level of competition in the radio space, in turn lowering the county’s exposure to the Show. Our  measure  of  competition is a spatial Herfindahl-Hirschman Index (HHI) in radio frequencies. While competition for a radio market in itself could be endogenous to the political preferences of a county, we build a measure of radio frequency competition based on accidental frequency overlaps in a county. The identifying assumption is that, conditional on the overall level of radio frequency competition in a county, the variation in radio frequency competition from accidental contour overlaps is not systematically correlated to variation in unobservables that affect election outcomes.

We first combine this measure of exposure to the Show with county-level election outcomes. We observe that counties with higher exposure to the Show have a systematically higher vote share for Donald Trump in the 2016 and 2020 U.S. presidential elections. We also find that the effect of the Show only becomes statistically and economically significant starting with the 2000 US presidential elections. This result mirrors two relevant facts around the Show and conservative politics in the US. First, while the Show has been airing since 1988, it did not attract much attention until the mid 90s \citep{jamieson08}.\footnote{In March 1994, Rush Limbaugh started raised red flags in the mainstream media and among Democrats with his announcement on air that Clinton White House confidant Vince Foster “was murdered.” A subsequent inquiry concluded that Foster had killed himself.} Second, the late 90s and early 2000s mark the rise of more populist groups within the Republican party (i.e. the Tea Party Movement\footnote{References to the Boston Tea Party were made during Tax Day protests since the early 1990s. An official website of the Tea Party movement declaring the Tea Party a nationwide movement, was launched in 2002}) who are both the result of the Show but also amplified its relevance and impact on the conservative electorate \citep{jamieson08}. This result is robust to alternative specifications.

Next, we combine our county-level, measure of exposure to the Show with individual survey data obtained from the Congressional Election Study (CES), to examine whether there are systematic differences in attitudes towards key social issues depending on exposure to the Show. Our examination reveals that individuals located in counties with higher exposure to the Show express stronger anti-abortion, anti- gay marriage, anti-immigration, anti-gun control and anti-environmental regulation attitudes. Interestingly, this is only the case for respondents who self-identify as Republicans, while the Rush-Limbaugh-exposure effect is largely absent among respondents self-identifying as Democrats.

This paper contributes to multiple strands of the literature. First, we relate to the broad economic literature on the effects of media in political outcomes in the U.S.\footnote{In this respect we also relate to the more extensive literature in economics on the media-politics nexus in other contexts such as newspapers and government responsiveness in India \citep{besley02}, the effect of free digital TV on election outcomes in Italy \citep{barone15} the effect of mobile Internet on political mobilisation in Africa \citep{manacorda20} social media and protests in Russia \citep{enikolopov20} and China \citep{qin21}, Internet and election outcomes in Germany \citep{falck11}, or Internet and trust in governments around the world \citep{guriev21}.}. Within this literature, studies have mainly focused on newspapers \citep[e.g.,][]{gerber09,gentzkow11b},  television\citep[e.g.,][]{Gentzkow06,DellaVigna07,gaeltta21,ash21} internet and social media \citep[e.g.,][]{Gentzkow11,campante13,melnikov21}. However, the effect of radio on U.S. politics has been largely understudied with most of the existing empirical work focusing on the historical perspective. \citet{stromberg04} showed how the expansion of radio in the 1920s led to more informed voters which in turn affected the allocation of relief spending under the New Deal program. More related to our paper, \citet{wang21} studies exposure to populist radio with a historical perspective, to the best of our knowledge, we are the first to study the effects of \textit{contemporary} talk radio on political outcomes. To our knowledge, \citet{barker99} and \citet{lee01} are the only other papers to empirically study the relationship between exposure to the Rush Limbaugh Show and voting outcomes. Using American National Election Studies (ANES) panel data from 1994 to 1996 \citet{barker99} found that respondents listening to the Rush Limbaugh show were more likely to vote for Republican candidates. However, the author explicitly acknowledges the challenges to causal inference in their setting. Our study does not only aim to address this identification problem highlighted by \citet{barker99} but also  analyses the effect of exposure to the Show on political attitudes, and political polarisation. 

With this in mind, we also join work on the determinants of political polarisation \citep[e.g.,][]{boxell17,draca21}. In particular, we follow \citet{boxell17} to construct measures of political polarisation from ANES survey responses. While their study focuses on the impact of internet exposure on political polarisation, we complement our analysis of the effect of the Rush Limbaugh Show on county level election outcomes, by further investigating the Show's  impact on political polarisation at the individual level further contributing to the literature on partisan media exposure and political polarisation \citep[e.g.,][]{sunstein09,levendusky13}.

While exposure to radio has been studied in other contexts, most of such work rely on the exogenous spatial variation in radio signal availability, as proposed by \citet{olken09} and applied in recent work such as \citet{enikolopov11}, \citet{adena15}, \citet{yanagizawa14} and \citet{blouin19}. Other branches of this literature exploit the variation in the staggered roll-out of programs \citep[e.g.,][]{DellaVigna07} or the position of a channel within the overall channel lineup \citep[e.g.,][]{martin17,ananyev21}. We make a methodological contribution to this literature by developing an alternative measure of exposure. The spatial HHI in radio frequencies developed in this paper, inspired by \citet{herfindahl50} and \citet{hirschman45}, can be more generally applied to facilitate empirical investigations in contexts where above methods cannot be applied. Conceptually, our identification strategy is in a similar spirit to \citet{barone15} who used a natural experiment that increased the number of free to view TV channels in Italy which decreased voters' exposure to the dominant, slanted, Berlusconi media.

The rest of this paper is organized as follows. In Section 2 we provide a brief introduction on talk radio in the U.S. and The Rush Limbaugh Show. In Section 3 we discuss the data. Section 4 discusses the empirical strategy and election results, while Section 5 demonstrates the individual-level effects. Section 6 concludes.

\section{Talk radio in the U.S. and The Rush Limbaugh Show}

Talk radio shows in the U.S. had its origins in the amplitude modulation (AM) radio space. AM broadcasting was the first method developed for making audio radio transmissions, and radio was the dominant method of broadcasting in the early 1970s when AM had around 75\% of the U.S. radio audience \cite{keith93}. This changed with the introduction of frequency modulation (FM) radio. Technological innovations in the 1970s and 1980s let to higher audio quality of FM radio and made it more suitable to the broadcasting of music and entertainment programs. With their lower audio fidelity, this resulted in a natural migration of AM radio away from music, and they became more prominently known for the specialized format of programs known as talk radio. 

Talk radio, a type of radio program that discusses and debates prominent social issues considered topical at the given point of time, were typically hosted by a prominent host, and the talk show itself was closely reflective of the host's own personality and perspectives. Early examples of this format trace back to the highly influential political talk radio show by Catholic priest, Father Charles Coughlin in the 1920s \citep{wang21}. While the AM listenership continued to decline with the rising competition from FM stations in the mid-20th century, one policy that changed the AM radio horizon was the repeal of the Federal Communication Commission (FCC)'s Fairness Doctrine in 1987. Prior to its repeal, the doctrine had required that talk radio shows present balanced information on topical issues, with the objective of exposing the audience to multiple viewpoints.  By 2011, it was estimated that there are close to 3,500 all-talk or all-news stations in the U.S. with the number to talk radio stations doubling just in the years between 2007 and 2011 alone \citep{berry11}.

One of the leading shows that capitalized early on from the repeal of the Fairness Doctrine  was The Rush Limbaugh Show.  The show, hosted by Rush Limbaugh himself, commenced in 1988 and delivered pro-conservative discussions and debates nationwide up. It first started as a local talk radio show in Sacramento in 1984, but expanded as a nationally syndicated talk radio show in 1988. The show did not attract much attention until March 1994, when Limbaugh starting spreading a rumor that a legal confidant of the Clinton White House, Vince Foster, was murdered. Although, a subsequent inquiry concluded that Foster had killed himself, and revealing Limbaugh's false claims, the event helped to boost the Show's nationwide popularity \citep{jamieson08}.

Until Rush Limbaugh's death in 2021, the show was delivered across approximately 585 radio stations and was aired for 3 hours during the day-time during week days. A weekend edition, featuring highlights of the week day edition, commenced in 2008. Since its inception, The Show has been widely acknowledged and promoting populist propaganda and controversial opinions.\footnote{See, for example, \textit{BBC}, ``Rush Limbaugh: How he used shock to reshape America," February 17, 2021. See also, \textit{The New York Times}, ``Talk radio is turning millions of Americans into conservatives," October 9, 2020.} 

Limbaugh draws his audience in and engages it in lengthy communication and discussion about the virtues of conservatism and the dangers inherent to liberalism and the ``liberal'' media. Thereby his show executes functions that were formerly identified with party leaders.  As with other ``conservative'' media (i.e. Fox News), the show reinforces a coherent set of rhetorical frames that empower their audiences to act as conservative opinion leaders, allowing Limbaugh in particular to mobilise party members for action, hold the Republican Party and its leaders accountable. In a world where the party identification of some individuals fluctuates with the political tides, listening to the Show may result in greater allegiance to the Republican Party. This generates a support basis that is more strongly aligned with conservative values, and more likely to support the Republican Party even when the Democrats present an appealing moderate or an independent candidate claims to be the genuine conservative in the election \citep{jamieson08}. 

Rush Limbaugh's persuasive communication can mobilise conservative voters and thereby impact the outcome of elections. For example, in November 1995, the Republican party won control of the House of Representatives for the first time in 40 years.
Republican leaders dubbed Limbaugh a ``majority creator'' and inducted him into the 104th Congress' rookie class as an honorary member. Tony Blankley, then republican leader Newt Gingrich's press secretary stated: ``After Newt, Rush was the single most important person in securing a Republican majority in the House of Representatives.'' \citep{jamieson08}. The Show also has substantial reach across the US population. Various reports\footnote{he State of the News Media, 2010, Pew Project for Excellence in Journalism (https://www.pewresearch.org/internet/2010/03/15/state-of-the-news-media-2010/), ``The Top Talk Radio Audiences,''
Talkers Magazine, March, 2011,
p. 22; \cite{berry11}} place Rush Limbaugh's number of weekly listeners anywhere between 13.5 and 15 million between 2003 and 2010.

Taken together, the Rush Limbaugh Show was one of the most popular, conservative talk-radio shows in the U.S between the mid 1990s until 2021. Limbaugh's audience is on average more politically involved  and he applied a rethoric that painted liberals, a ``cultural elite'' and democrats as ``enemies'' and a ``threat''  \citep{jamieson08}. This rethoric around enemies of America was a strong unifying force to mobilise his listeners during elections. In addition, it moved his already predominately conservative listeners to even more conservative positions, thereby increasing overall political polarisation in the United States. 

\section{Data}

\subsection{Data on exposure to The Rush Limbaugh Show}

To identify each county's exposure to The Rush Limbaugh Show, we first obtain the list of radio stations that delivers the Show across the U.S., from the Show's official website.\footnote{\href{www.rushlimbaughshow.com}{https://www.rushlimbaugh.com/}} We identify 585 U.S.-based stations delivering the show, 347 of which are on the AM frequency.\footnote{Of the remaining stations, 112 stations are on the FM frequency, while 120 are live streaming channels.} Our specific focus is on \textit{AM stations} delivering the show which have, since historically, specialized in the delivery of talk shows. As discussed, AM stations are highly susceptible to interference compared to FM stations and have lower audio fidelity, making them less suitable to the delivery of entertainment and musical programs and more suited for talk radio. 

\begin{figure}[htpb!] 
\caption {AM and FM contours}
\label{fig:contours}
\begin{subfigure}[b]{\textwidth}
\begin{center}
    \includegraphics[width=12cm,height=16cm,keepaspectratio]{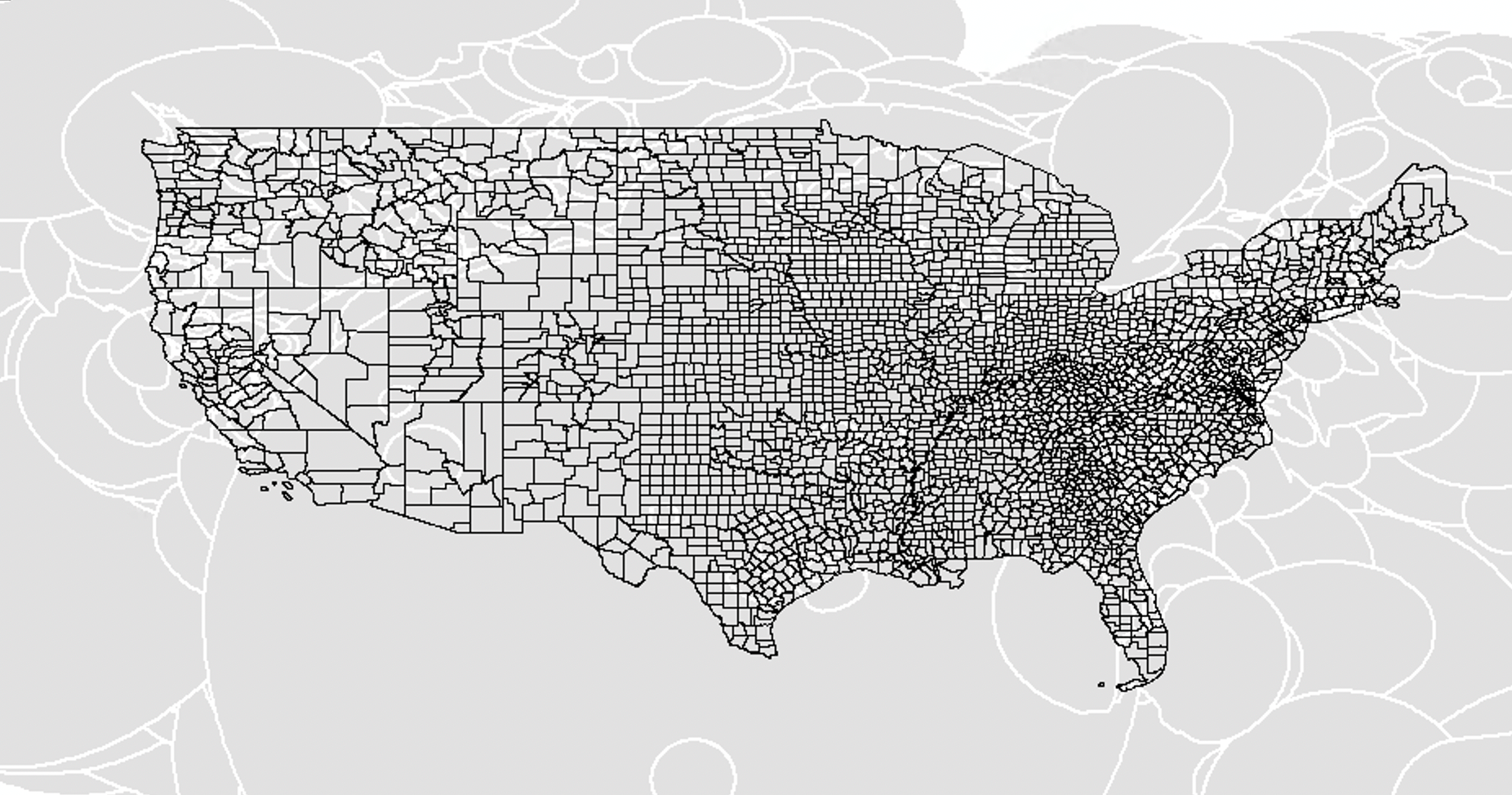}
    \caption{Spatial distribution of AM contours}
    \end{center}
    \end{subfigure}
    \hspace{10mm}
    \begin{subfigure}[b]{\textwidth}
    \begin{center}
   \includegraphics[width=12cm,height=16cm,keepaspectratio]{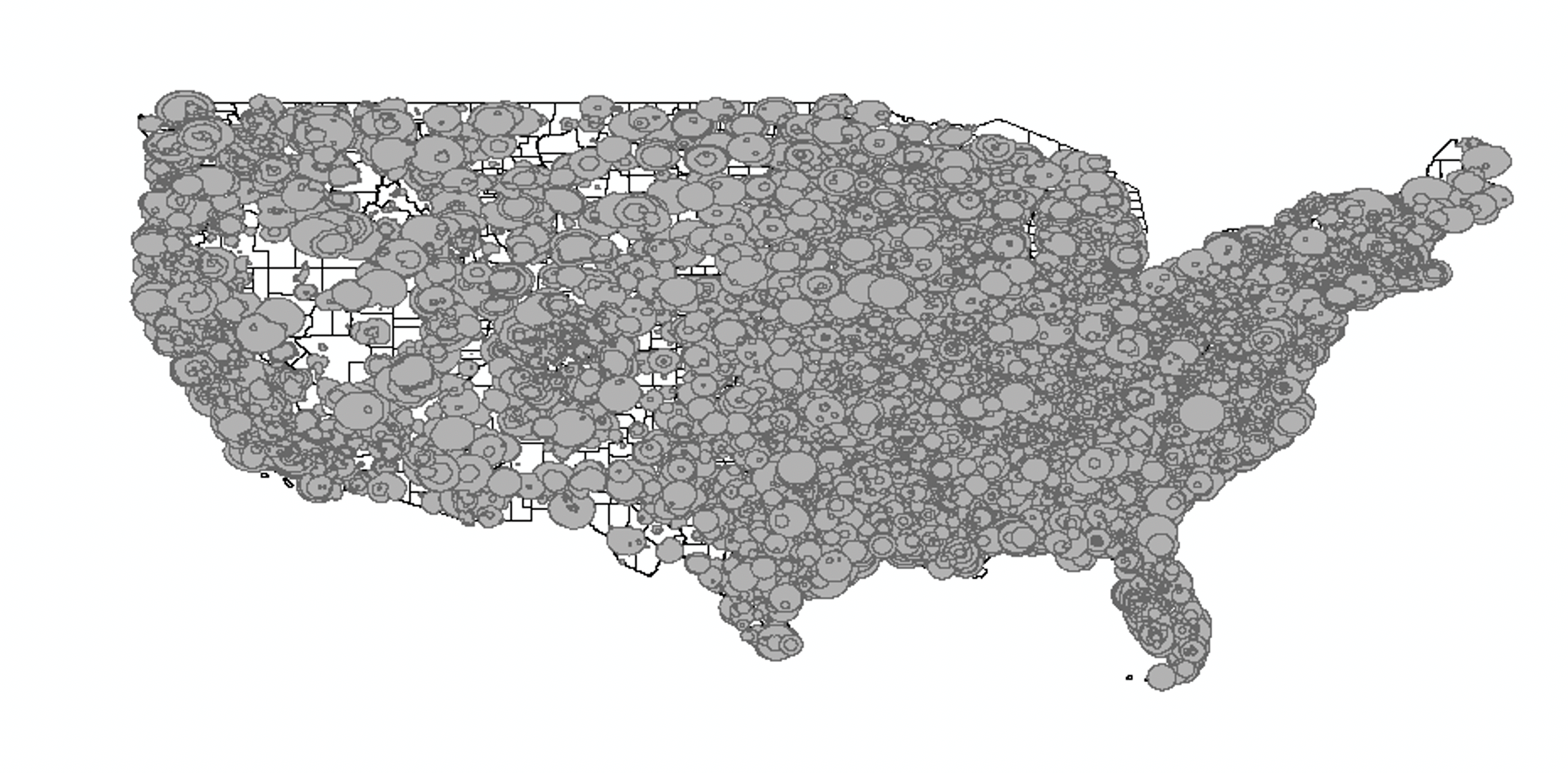}
    \caption{Spatial distribution of AM contours}
    \hspace{30px}
     \end{center}
    \end{subfigure}
   \captionsetup{width=\linewidth}
    \end{figure}

Next, we obtain data on AM contours in the U.S. from the Federal Communications Commission (FCC). Panel (a) in Figure \ref{fig:contours} shows the spatial distribution of these AM contours across the U.S. We observe that AM contours are broad in their coverage and are spatially distributed in a manner that covers the entire U.S. It is also important to distinguish \textit{AM stations} from \textit{AM contours} -- each AM station possesses multiple contours at different levels of electric field strength intensity, as measured by millivolts per meter (mV/m). These contours also receive varying levels of protection against interruptions from adjacent and co-channels, depending on whether it is a daytime or nighttime contour.

For the purpose of our analysis, we first identify within the FCC data, the set of AM contours belonging to the AM stations listed on The Rush Limbaugh Show's website. We identify 1,388 contours spread across the country that belong to AM stations airing the show. By overlapping these AM contours with county boundaries, we identify the number of AM contours broadcasting the Show in each county.\footnote{Since The Rush Limbaugh Show was typically aired during the daytime, we only retain the daytime groundwave contours belonging to each AM station.} Figure \ref{fig:contours} shows the dispersion of AM contours airing the Show across the U.S. We observe that there is a high concentration of The Show along the east and west coasts and the mid-west.

\begin{figure}[htpb!] 
\caption{County-wide dispersion of AM contours broadcasting The Rush Limbaugh Show}
    \begin{center}
   \includegraphics[width=12cm,height=16cm,keepaspectratio]{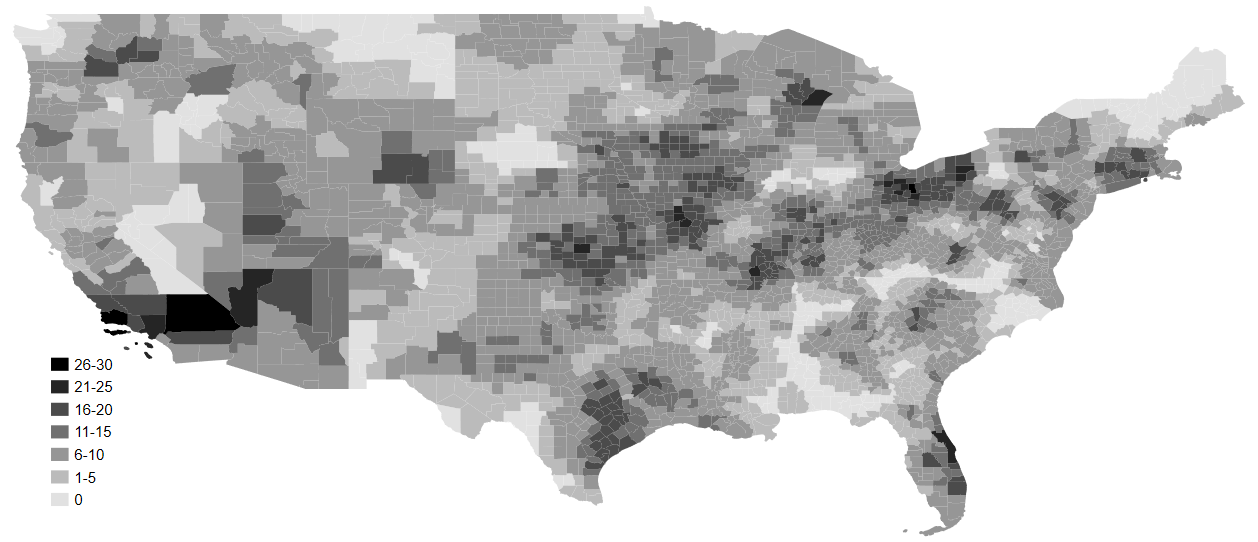}
    \hspace{30px}
     \end{center}
  \captionsetup{width=\linewidth}
    \caption*{\footnotesize{\textit{Source: } Figure shows the number of daytime AM contours broadcasting The Rush Limbaugh Show in each county.}}    
    \end{figure}
    
\subsection{Data on FM contours}
\label{sec:accidental}
\begin{figure}[htpb!] 
\caption {Distribution of FM contours}
\label{fig:FM_contours}
\begin{subfigure}[b]{\textwidth}
\begin{center}
    \includegraphics[width=13cm,height=18cm,keepaspectratio]{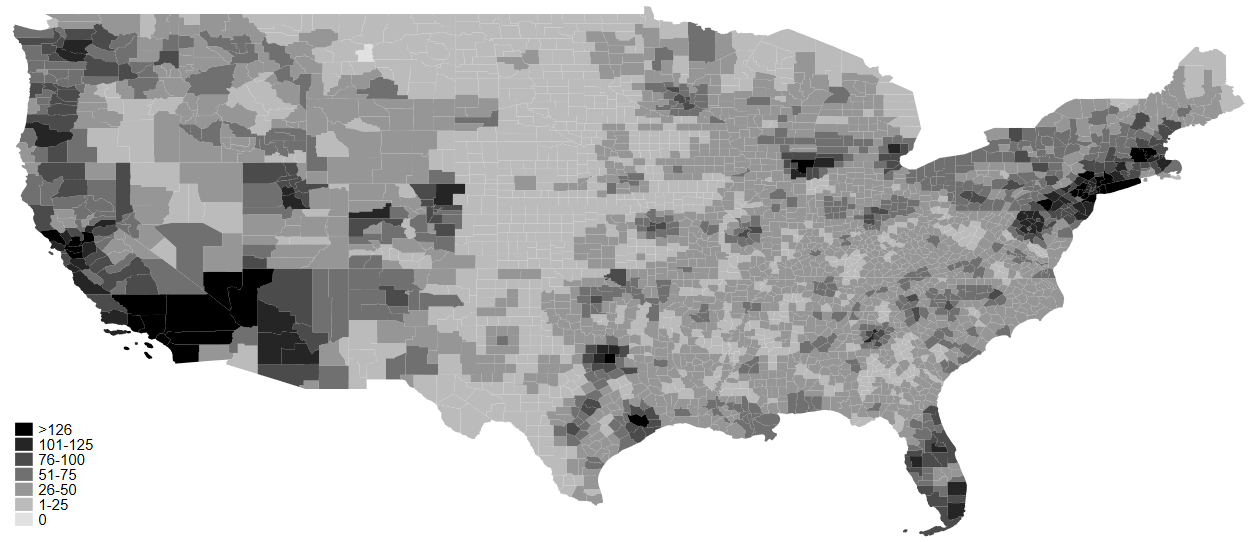}
    \caption{County-wide dispersion of all FM contours}
    \end{center}
    \end{subfigure}
    \hspace{10mm}
    \begin{subfigure}[b]{\textwidth}
    \begin{center}
   \includegraphics[width=13cm,height=18cm,keepaspectratio]{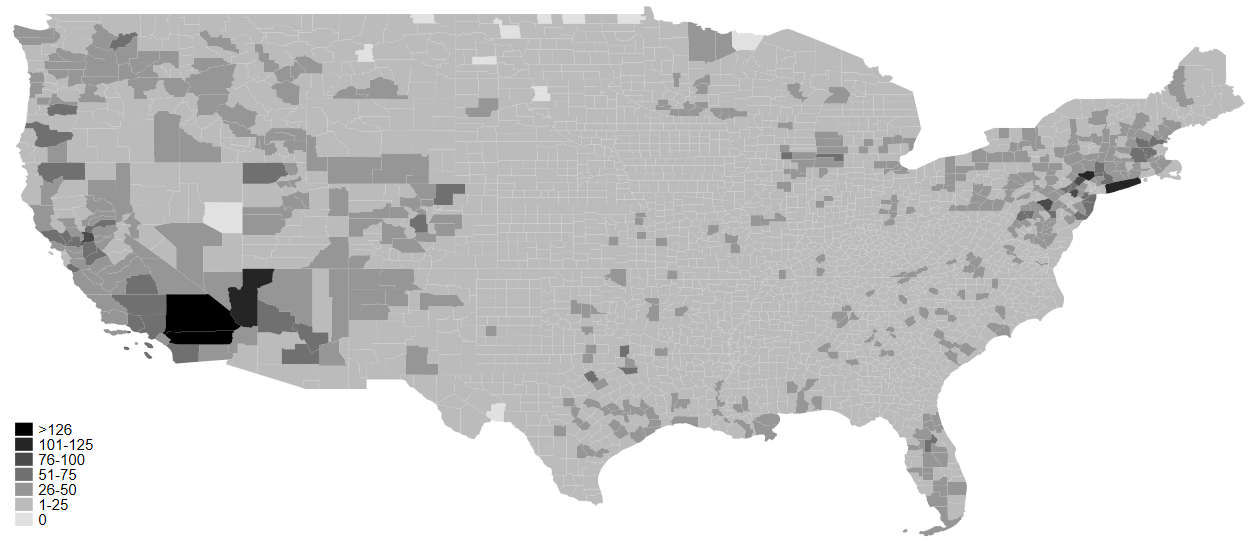}
    \caption{County-wide dispersion of ``accidental'' FM contours}
    \hspace{30px}
     \end{center}
    \end{subfigure}
     \begin{subfigure}[b]{\textwidth}
    \begin{center}
   \includegraphics[width=13cm,height=8cm,keepaspectratio]{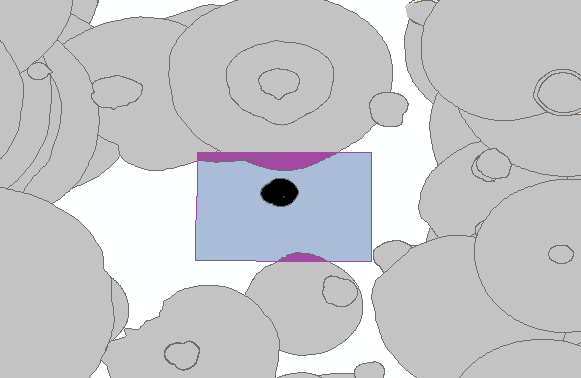}
    \caption{Identifying accidental FM contours}
    \hspace{30px}
     \end{center}
    \end{subfigure}
   \captionsetup{width=\linewidth}
    \end{figure}

Our empirical strategy exploits the competition faced by the Show from FM contours. To generate the indicator of competition, we first obtain data on the spatial distribution of FM contours from the FCC. Panel (b) in Figure \ref{fig:contours} shows the dispersion of FM contours across the U.S. We observe that FM contours are narrower and more specific in their coverage when compared to AM contours. As with AM contours, we overlap the FM contours with county boundaries to identify the FM exposure in each county. In its simplest form, this overlap can identify the number of FM contours received by each county, as indicated in Panel (a) of Figure \ref{fig:FM_contours}.

For the purpose of the identification strategy pursued in this paper, we go a step beyond this `naive' indicator of FM coverage and distinguish between the ``intentional'' vs ``accidental'' FM coverage in each county. To understand the intuition behind this distinction, consider the setting in Panel (c) of Figure \ref{fig:FM_contours}. Here, the rectangular blue polygon represents Baca county in the state of Colorado. The circular shapes are FM contours. The contour depicted in black is entirely encapsulated within the county borders, and therefore it seems reasonable to assume that this FM contour was intentionally placed for reception by this county. We consider such FM coverage as ``intentional" FM coverage. The grey contours that do not overlap with county borders do not contribute towards the FM exposure of Baca county. The purple polygons represent overlaps between the county boundaries and peripheral FM contours which, although not fully covering the county, do provide a marginal level FM exposure. Considering their peripheral location and marginal coverage of county area, it seems reasonable to assume that these contours were not specifically placed targeting Baca county, although the county does ``accidentally'' receive FM coverage from these contours. We consider such FM coverage ``accidental'' coverage for this county.

Accordingly, for the purpose of our empirical exercise we consider all contours that either (a) cover the entire county or (b) that are completely located within the county as ``intentional'' FM coverage. Of the remaining contours, those with a coverage area more than the median size of the overlapping polygons are identified as ``intentional'' FM coverage. All contours where the size of the overlapping polygon is less than the median value of all overlapping polygons is identified as ``accidental'' FM coverage for the given county.\footnote{Note that the definition of ``intentional'' and ``accidental'' FM coverage is county-specific. An FM contour which is accidental for county X may or may not be accidental for county Y, depending on the size of the overlapping polygon.} Panel (b) of Figure \ref{fig:FM_contours} provides the distribution of accidental FM contours across the U.S. It is important to note that this represents a subset of the total number of FM contours depicted in Panel (a) of Figure \ref{fig:FM_contours}.

\subsection{Data on election outcomes}
We obtain county-level data on election outcomes in the U.S. from the Atlas of U.S. Presidential Elections. We calculate the Republican vote share for each presidential election for each county, going back to 1968. 

\subsection{Data on individual political views and policy preferences}
We use individual level survey data from the 2016 Cooperative Elections Study (CES), to identify individual attitudes on key social issues. This survey consists of 64,600 respondents from across the U.S. Importantly, for each respondent, the survey provides their geo-location (i.e., county) which allows us to match the county-level indicators of exposure to the Show to each individual. 

We focus on five questions in this survey that capture the respondents' attitudes towards abortion, gay marriage, granting legal status to immigrants, deporting illegal immigrants, gun control and environmental regulations. The answers to these question can either be binary (Support/Do not support) or hedonic (Strongly support/Somewhat support/Neither support or oppose/Somewhat oppose/Strongly oppose) responses. We convert these responses to binary format, by generating indicators which assume a value of 1 if the respondent supports/strongly supports a statement, and zero otherwise.











    
%

\section{Empirical strategy}

\subsection{Deriving indicators of county-level competition}

The degree to which listeners within a particular county are exposed to the show depends not only on the number of contours broadcasting the Show, but also on the number of alternative contours (i.e. that do not broadcast the Show) received by the county. For example, the exposure to the Show will be higher in counties where the only radio station received is one that broadcasts the show, as opposed to another county where there are many alternative channels. Accordingly, we focus on the radio space in each county as a market where multiple stations are competing with each other for listener's attention. We consider FM stations, which primarily deliver entertainment and musical programs, as the key competitor to AM stations delivering the Show. Our hypothesis is that a higher level of `alternative' channels increases the level of competition in the radio space, in turn lowering the county's exposure to the Show.

Our measure of competition in the radio market is inspired by the Herfindahl-Hirschman Index ($HHI$) \citep{herfindahl50, hirschman45} of market competition.
To calculate the $HHI$, we first overlap AM contours (for stations delivering the Rush Limbaugh show) and FM contours with county boundaries. Based on this overlap, we identify approximately 1.2 million unique intersecting polygons $p$, and the number of AM and FM contours belonging to each such unique polygon. We then calculate the relative share occupied by AM and FM stations within each unique polygon and calculate the $HHI$ as per the standard $HHI$ equation in Equation \ref{eq:HHI_all} below.

\begin{equation}
\label{eq:HHI_all}
   HHI all_{i}= \sum_{p=1}^{N} RL Share^2_{p,i} + \sum_{p=1}^{N} FM Share^2_{p,i}
 \end{equation}
 
where $RL Share^2_{p,i}$ is the squared market (geographic) share of all AM stations delivering the Rush Limbaugh show, for the unique intersecting polygon $p$ in county $i$. Likewise, $FM Share^2_{p,i}$ is the squared market share of all FM stations received by the  unique intersecting polygon $p$ in county $i$. The $HHI$ is typically valued between 0 and 1, with higher values signalling less competition (more monopoly power). 

One concern related to this $HHI$ however is that FM stations and their contours are likely strategically placed to maximise coverage - a more populous county would be covered by more FM contours as opposed to a less populous county. Therefore, an identification strategy that simply considers the `naive' AM-FM competition level within each county, as demonstrated in Equation \ref{eq:HHI_all} above, will likely suffer from endogeneity bias.
 
We observe, however, that when planning the ``intentional'' location of an FM station, some surrounding counties might receive FM coverage ``accidentally''. Exploiting such accidental FM coverage will allow us to filter out the quasi-random variation in the $HHI$ which will in turn enable the causal interpretation of our estimates. As already discussed in Section \ref{sec:accidental}, we define an FM station as `accidental' from the perspective of a county, if the overlapping area between the FM contour and the county is less than the median value of all such overlapping areas for the whole sample.\footnote{Our baseline estimates are robust to alternate cutoffs in defining the accidental nature of FM contours, as presented in Table \ref{tab:election_alt}.} It is important to note that an FM station identified as accidental from the perspective of one county may or may not be accidental for another county, depending on the area covered by each FM contour within each county. We then recalculate the $HHI$ considering the competition posed only by these accidental FM contours, using Equation \ref{eq:HHI_acc}. 
 
\begin{equation}
\label{eq:HHI_acc}
   HHIacc_{i}= \sum_{p==1}^N RL Share^2_{p,i} + \sum_{p==1}^N  AccFM Share^2_{p,i} 
 \end{equation}
 
\begin{figure}[htpb!] 
\caption {HHI}
\label{fig:HHI}
\begin{subfigure}[b]{\textwidth}
\begin{center}
 \includegraphics[width=15cm,height=18cm,keepaspectratio]{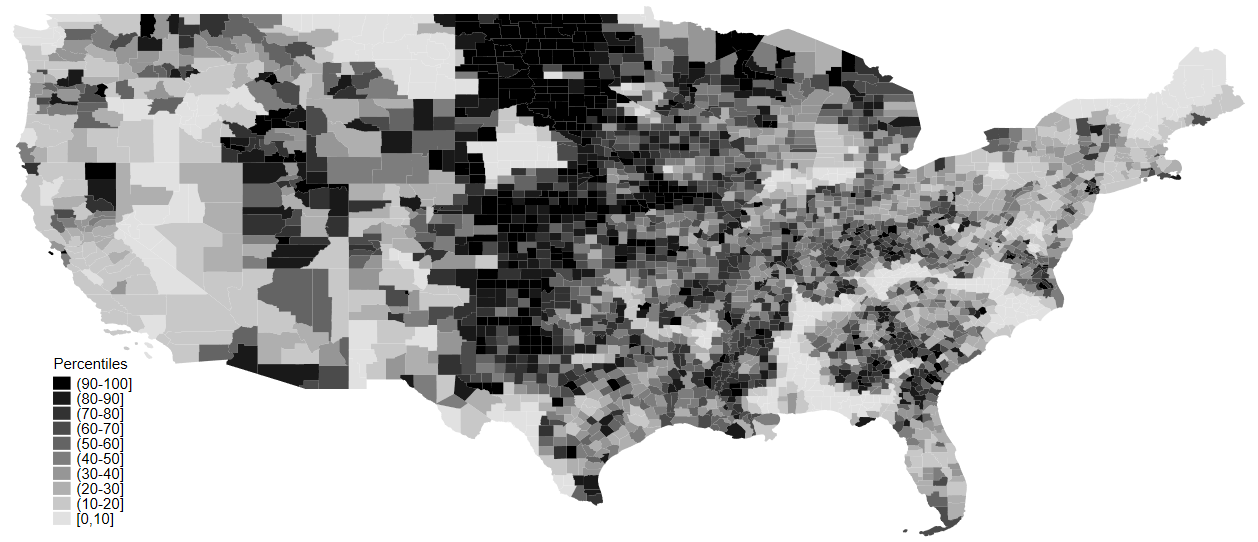}
    \caption{HHI based on all FM contours}
    \end{center}
    \end{subfigure}
    \hspace{20mm}
    \begin{subfigure}[b]{\textwidth}
    \begin{center}
   \includegraphics[width=15cm,height=18cm,keepaspectratio]{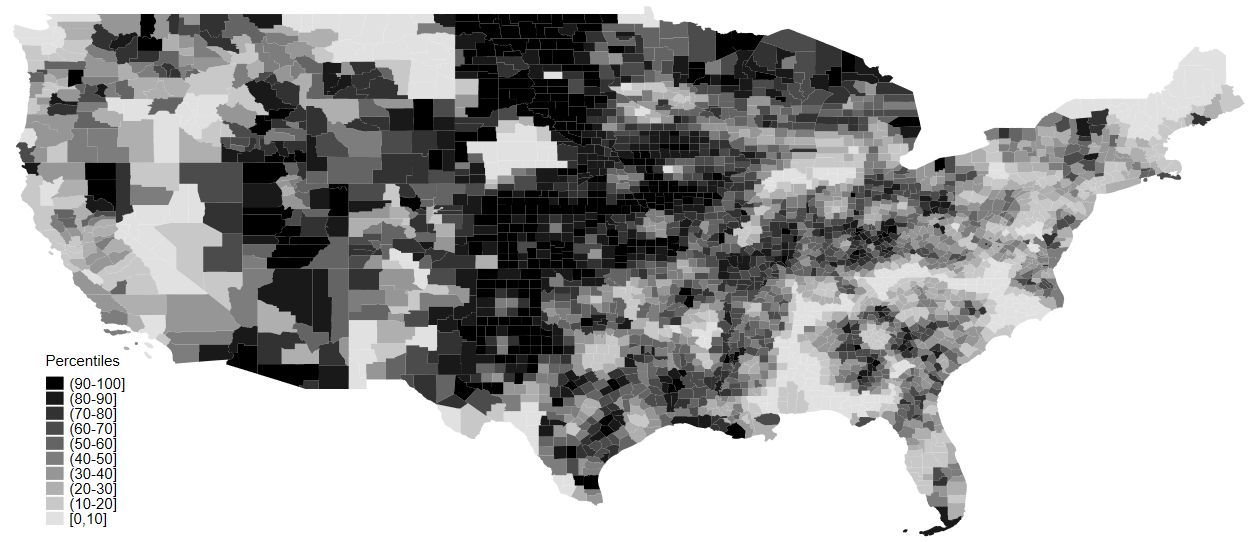}
    \caption{HHI based on accidental FM contours}
    \hspace{30px}
     \end{center}
    \end{subfigure}
   \captionsetup{width=\linewidth}
    \caption*{\footnotesize  \textit{Notes:} Panels (a) and (b) show the distribution of the HHI index based on all and accidental FM contours, based on Equations \ref{eq:HHI_all} and \ref{eq:HHI_acc}, respectively. An FM contour is identified as accidental for a given county if if the overlapping area between the FM contour and the county is less than the median value of all such overlapping areas for the whole sample. }    
    \end{figure}

Here, $AccFM Share^2_{p,i}$ is the squared market share of accidental FM stations received by the  unique intersecting polygon $p$ in county $i$. Again, this $HHIacc$ is typically valued between 0 and 1, but is lower than $HHIall$ as it exploits only a subset of the competition incorporated in the latter. Panels (a) and (b) in Figure \ref{fig:HHI} display the percentile distribution of $HHIall$ and $HHIacc$, respectively.

CAVEATS: It is important to note that in these estimates is that we focus only on the AM delivery of the Show. Recently some FM stations have also started to air the Show. This means that our competition indices underestimate the market presence of the Show, and our estimates are therefore likely to be biased downwards.\footnote{The absence of a precise matching identifier in the FM contours data set and Rush-Limbaugh-delivering-FM-contours data set precludes us from quantifying this bias.} Moreover, as of now we are only considering the competition to the Show from FM contours. The current estimates do not account for competition arising from other AM contours that do not broadcast the Show. 

\begin{figure}[htpb!]
\caption{Correlation between $HHIacc$ and county level characteristics}
\label{fig:overtime}
\begin{center}
	\includegraphics[width=16cm,height=18cm,keepaspectratio]{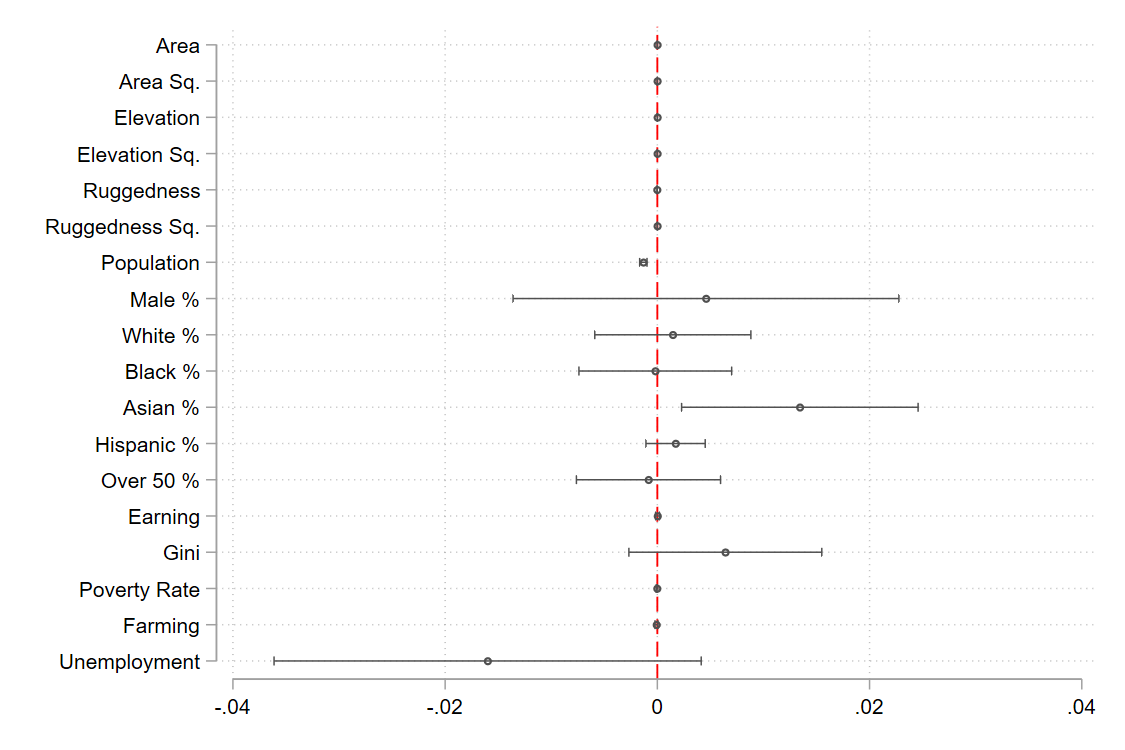}
	\end{center}
\captionsetup{width=\linewidth}
\caption*{\footnotesize{\textit{Notes:} Dependent variable is $HHIacc_{i}$. Figure shows the correlation between $HHIacc$ and a range of geographic and demographic variables in each county. Dots show the point estimates while the vertical lines depict the 95\% confidence intervals, based on standard errors clustered at the State level.} }  
\end{figure}

\begin{table}[htpb!]
\begin{center}
\caption{Descriptive Statistics for Key Variables}
\label{tab:sumstats}
\resizebox{1\textwidth}{!}{
\begin{tabular}{lccccc} \toprule
 
&No. of &Mean &Standard & Minimum & Maximum\\ 
 & Observations & & Deviation\\
\midrule
$HHIall$ & 1,910 & 0.0110 & 0.0129 & 0 & 0.2646 \\
$HHIacc$ & 1,910  & 0.0024 & 0.0042 & 0 & 0.0509\\
\\
$Rep$ $Vote$ $Share$ $2016$ &  1,906 & 0.6021 & 0.1507& 0.0841& 0.8948\\
$Rep$ $Vote$ $Share$ $2020$ &  1,817 & 0.6310 & 0.1776& 0& 1\\

\\
$Support$ $Abortion$ & 64,538 & 0.6168 & 0.4862&0 &1 \\
$Support$ $Gay$ $Marriage$& 64,125 & 0.6506 & 0.4768& 0&1 \\
$Support$ $Legal$ $Status$ $to$ $Immigrants$ & 64,600 & 0.5601 & 0.4964& 0& 1\\
$Support$ $Deporting$ $Immigrants$ &64,600  & 0.4063 &0.4912 & 0& 1\\
$Support$ $Gun$ $Control$ & 64,223  & 0.6225 &0.4848 &0 & 1\\
$Support$ $Environmental$ $Regulation$ & 64,547 & 0.5836 & 0.4930& 0& 1\\
\bottomrule
\end{tabular}}
\end{center}
\end{table}

\subsection{Empirical strategy}
To estimate the effect of exposure to the Show on electoral outcomes, we estimate the following county-level specification.

\begin{equation}
\label{eq:main}
    RepVoteShare_{c,s} = \beta_{1} HHIall_{c,s} + \beta_{2}HHIacc_{c,s} + \beta_{3}X_{c,s} + \mathbf{FE_{s}}+ \epsilon_{c,s}
\end{equation}
where $RepVoteShare_{c,s}$ is the Republican vote share in county $c$ of state $s$. $HHIall_{c,s}$ is the Herfindahl-Hirschman Index of competition faced by AM contours delivering The Rush Limbaugh Show from all (i.e. intentional and accidental) FM contours in county $c$ of state $s$, calculated as per Equation \ref{eq:HHI_all}. $HHIacc_{c,s}$ is the competition faced by AM contours delivering the Show from purely accidental FM contours in county $c$ of state $s$, calculated as per Equation \ref{eq:HHI_acc}. \textbf{X} is a vector of geographic and demographic controls at the county level. $FE_{s}$ is a vector of state fixed effects that accounts for any state-level unobservables. The coefficient of interest, $\beta_{2}$, identifies the effect of competition faced by the Show from accidental FM coverage on the Republican vote share, conditional on the effect of total competition captured by $\beta_{1}$. Considering the dominant pro-Republican agenda promoted by the Show, we expect $\beta_{2}$ to be positive. This approach estimates and intention-to-treat (ITT) effect and the estimated effects are likely to be lower that the true effect.

To begin with, we focus on the two most recent US presidential elections. Panels A and B in Table \ref{tab:election} show the estimates for the 2016 and 2020 elections, respectively. In Column (1) we show the unconditional effect of $HHIacc$ with no controls. As expected, the coefficient is positive and highly statistically significant, meaning that a high level of HHI in a given county (equivalent to lower competition faced by the Show) increases the Republican vote share in the same county. In terms of economic significance, a one standard deviation increase in $HHIacc$ increases the republican vote share by approximately 2.5 and 3 percentage points in panles A and B respectively.  This effect holds when controlling for State fixed effects in Column (2). In Column (3) we include $HHIall$ as a control variable, so that The coefficient on $HHIacc$ is the effect of purely accidental competition faced by the Show on the Republican vote share, conditional on the competition derived from all FM stations. In Columns (4), (5) and (6) we additionally control for a rich set of geographic, demographic controls and historical voting patterns, respectively, and observe that the estimated effect is robust to such inclusions.

\begin{table}[htpb!]
\small
\caption{Effect of exposure to The Rush Limbaugh Show on Republican vote share}
\label{tab:election}
\begin{center}
\small
\resizebox{\textwidth}{!}{
\begin{tabular}{lcccccc} \toprule
 & (1) & (2) & (3) & (4) & (5) & (6) \\
& \multicolumn{6}{c}{\textbf{A: Dependent Variable:} \textit{2016} $Republican$ $Vote$ $Share_{c,s}$}\\ \midrule
 &  &  &  &  &  &  \\
$HHIacc_{c,s}$ &   6.0689*** &  5.3216*** &  7.5357*** & 6.9962*** & 1.9599*** & 1.7395*** \\
 & (1.2781) & (1.0603) & (1.2275) & (1.2315) & (0.7107) & (0.6485) \\
$HHIall_{c,s}$ &  &  &   -1.5915*** & -1.6400** & -0.3953 & -0.3564 \\
 &  &  & (0.5096) & (0.6636) & (0.3792) & (0.4091) \\
 &  &  &  &  &  &  \\
Observations & 1,901 & 1,901 & 1,901 & 1,885 & 1,885 & 1,885\\

\midrule

& \multicolumn{6}{c}{\textbf{B: Dependent Variable:} \textit{2020} $Republican$ $Vote$ $Share_{c,s}$}\\ \midrule
 &  &  &  &  &  &  \\
$HHIacc_{c,s}$ &  7.3501*** & 5.8952*** &  8.3870*** & 7.7596*** &  2.8531*** & 2.6280*** \\
 & (1.4274) & (1.1901) & (1.4437) & ( 1.4100) & (0.9835) & (0.9337) \\
$HHIall_{c,s}$ &  &  & -1.7872***  &  -1.9237** & -0.7907* & -0.7576 \\
 &  &  & (0.5797) & (0.7397) & (0.4650) & (0.4768) \\
 &  &  &  &  &  &  \\
Observations &   1,812 & 1,812 & 1,812 &  1,796 & 1,796 & 1,796 \\
\midrule
State FE & NO & YES & YES & YES & YES & YES \\
Geographic Controls & NO & NO & NO & YES & YES & YES \\
Demographic Controls & NO & NO & NO & NO & YES & YES \\
Avg. Rep. Share 1968-1984 & NO & NO & NO & NO & NO & YES \\ \bottomrule
\multicolumn{7}{p{1.1\textwidth}}{\scriptsize{\textit{Notes:} The dependent variable in Panels A and B is the Republican vote share in the 2016 and 2020 presidential elections, respectively. Geographic controls include county area, elevation, ruggedness and their respective squared terms. Demographic controls include log of total population, population shares for male, black, Asian, Hispanic and above 50 year categories, log of median earnings, Gini coefficient, poverty rate, unemployment and farming area. Standard errors, clustered at the State level, are in parenthesis. *** p$<$0.01, ** p$<$0.05, * p$<$0.1}} \\
\end{tabular}}
\end{center}
\end{table}

\begin{figure}[htpb!]
\caption{Effect of exposure to The Rush Limbaugh Show on Republican vote share over time}
\label{fig:overtime}
\begin{center}
	\includegraphics[width=12cm,height=14cm,keepaspectratio]{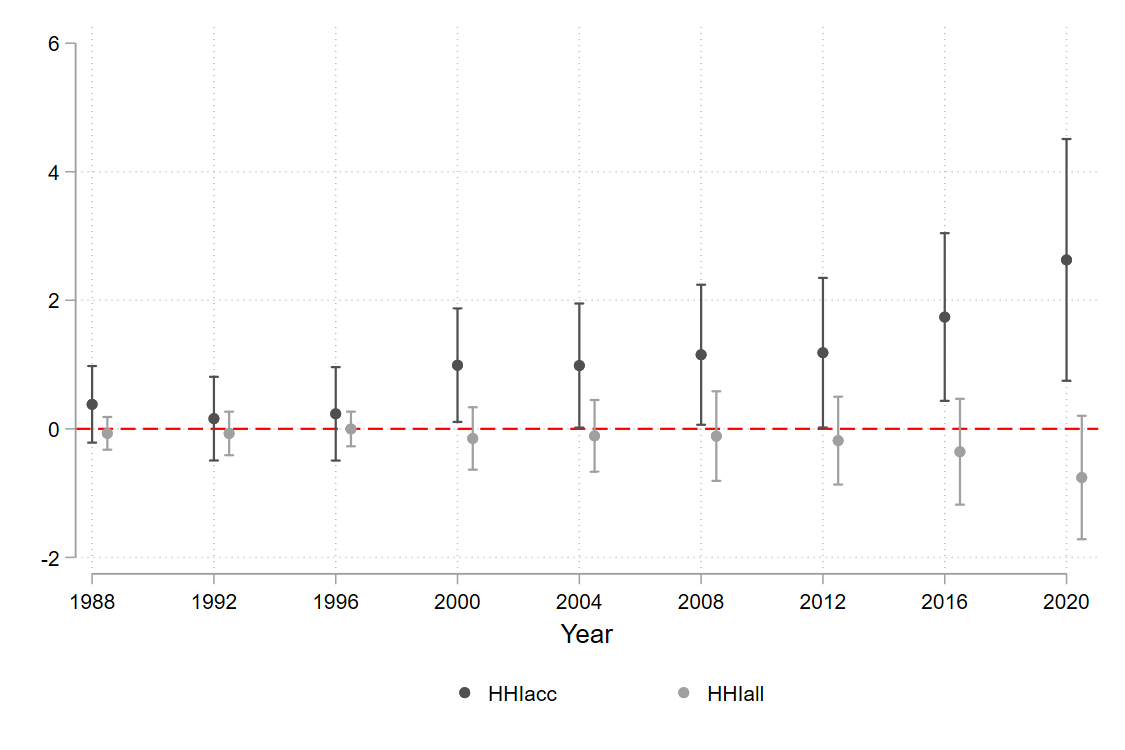}
	\end{center}
\captionsetup{width=\linewidth}
\caption*{\footnotesize{\textit{Notes:} Figure shows the effect of $HHIacc_{c,s}$ and $HHIall_{c,s}$ on Republican vote share over time. Dots show the point estimates while the vertical lines depict the 95\% confidence intervals, based on standard errors clustered at the State level. Estimates for each election year represent a separate regression estimate. All estimates include geographic (county area, elevation, ruggedness and their respective squared terms) and demographic (log of total population, population shares for male, black, Asian, Hispanic and above 50 year categories, log of median earnings, Gini coefficient, poverty rate, unemployment and farming area) controls. Estimates also control for the average Republican vote share over 1968-1984.} }  
\end{figure}

Next, we complement these estimates with an examination of the effect of the Show on the Republican vote share since its inception in 1988. In Figure \ref{fig:overtime}, observe that the effects are more precisely observed from the early 2000's, and are increasing in magnitude up to 2020. As mentioned before, prior to the mid 1990s, the Show did not attract a lot of nation-wide attention. His audience started to grew around the events of the Vince Foster story in March 1994 and following the Republicans win of the House Representatives elections in November 1994. The Shows popularity saw a further boost with the advent of the Obama administration in 2009. In addition, the Tea Party movement, a more fiscally conservative faction within the Republican party was launched. Members of this movement were even more receptive to Rush Limbaugh's rhetoric and also acted as an amplifier of his messaging to the broader conservative electorate.  

\subsection{Robustness checks}  
Now we examine the robustness of these baseline estimates to alternative specifications.

Given the spatially clustered nature of the exposure to the show, we first examine the robustness of the baseline estimates when adjusting for standard errors accounting for spatial correlations, as per \citet{conley99}. In Figure \ref{fig:overtime_conley}, we show that the results are robust to adjusting for spatial correlations for up to 500km, in 100km intervals. Moreover, in Table \ref{tab:election_spatiallags}, we show that the estimates are robust to alternative models, i.e. Spatial autoregressive and Spatial Durbin models, based on the contiguity matrix of adjacency.

In Table \ref{tab:election_placebo}, we show that the measure of exposure to the Show did not have an effect on the election outcomes in the period prior to the inception of the Show, which confirms the validity of our finding on the effects of the exposure to the Show on election outcomes.

Recall that in the baseline estimates, we considered an FM contour to be ``accidental'' if the size of the overlap between the contour and county area is less than the median value of all such overlapping areas. We now consider alternative cutoffs for this definition. Accordingly, in Column (1) of Table \ref{tab:election_alt}, an FM contour is considered accidental if the overlap area is less than the 25th percentile of all such overlapping areas, while in Column (2) this cutoff is based on the average value of overlapping areas. We observe that relaxing this definition does not affect the baseline estimates qualitatively or quantitatively.

Next we consider whether the baseline estimates are driven by accidental contours in uninhabited areas. However, the empirical barrier for such an estimation is the absence of spatially granular population data for each of these overlapping polygons. To overcome this problem, we utilize data on nighttime lights. By overlapping geo-coded nighttime light data with the accidental FM contours, we are able to calculate the amount of nighttime light under each such accidental contour. Using nighttime lights as a proxy for population, we then exclude all accidental FM contours where the nighttime light is less than the 10th percentile of the total nighttime light distribution. In Column (3) of Table \ref{tab:election_alt} we observe that the baseline result does not change significantly when excluding accidental contours in uninhabited areas.

\newpage
\section{Effects on individual attitudes}
Next we focus on how exposure to the Show affects individual political views. For this purpose, we use annual survey data from the Cooperative Election Study (CES, formerly the CCES), covering approximately 500,000 respondents over the years 2006-2020. Particularly relevant for our purpose, the CES provides the geo-location of each respondent (i.e., the county), which we allows us to link them to our exposure measure. We then define the following specification. 

\begin{equation}
\label{eq:survey}
    Outcome_{i,c,s,y} = \gamma_{1} HHIall_{c,s} + \gamma_{2}HHIacc_{c,s} + \beta_{1}X_{i,c,s}  + \beta_{2}Z_{c,s}+ \mathbf{FE_{s}} + \mathbf{FE_{y}} + \epsilon_{i,c,s}
\end{equation}

where $Outcome_{i,c,s,y}$ is a binary indicator on the political views and policy preferences of respondent $i$ residing in county $c$ of state $s$, in year $y$. As before, $HHIall_{c,s}$ and $HHIacc_{c,s}$ represent county level exposure to The Rush Limbaugh Show, based on all FM contours and accidental FM contours, respectively. \textbf{X} is a vector of individual level controls, while \textbf{Z} is a vector of county level (geographic and demographic) controls. $FE_{s}$ is a vector of state fixed effects that accounts for any state-level unobservables, while $FE_{y}$ is a vector of year fixed effects that absorbs any time-varying, year-specific unobservables. The coefficient of interest, $\gamma_{2}$ identifies the effect of county-level (accidental) exposure to the Show on political views of individuals belonging to the same county $c$. Again, this approach estimates an intention-to-treat effect.

We first examine individual political views. We define binary indicators identifying respondents' political views based on answers to the question, ``How would you define your political views?''. This question yields a set of hedonic answers ranging from ``Very Conservative", ``Conservative", ``Moderate'', ``Liberal'' or ``Very Liberal''. We use this information to define three binary indicators - equalling to one if the respondent declared their political views as (a) ``Strong Conservative'', (b) ``Conservative'' and (c) ``Moderate'', and zero otherwise - and use these as the dependent variables in Equation \ref{eq:survey} above. 

Rush Limbaugh's regular audience consists largely of Republicans. It is therefore very likely that Democrats do not really listen to the stations airing the show and are therefore not exposed (``treated''). Even if they listen to the Show, \citet{mutz01} study on Americans exposure to dissimilar political views showed that Democrats are more likely than Republicans to find that the views expressed in talk-radio shows are in disagreement with their own. To examine the potential heterogenous effects of exposure to the show on respondents with different party preferences, we build an interaction term between our HHI measures, $HHIall_{c,s}$ and $HHIacc_{c,s}$, and $Rep_i$, a dummy that switches to one if the respondent voted for the Republican party in the previous elections and zero otherwise.



\subsection{CCES Estimates}

Table \ref{tab:polar} and Figure \ref{fig:attitude} present the results of the effect of exposure to the Show on individual political attitudes. Ignoring the respondents party preference, we do not find that exposure to the Show leads to more conservative attitudes (columns 1 and 3) but respondents tend to consider themselves as more `Moderate'. However, once we include the interaction term that indicates if the respondent has voted for the Republican party in the previous elections, we find some interesting patterns. Self-reported Republicans tend to have stronger conservative political views (columns 2 and 4) and also less moderate (column 6). In contrast, Democrats (which are captured by $HHIacc_{c,s}$\footnote{Note, we excluded respondents that voted for Independent candidates in the previous election.}) located in counties with high exposure to the show are less likely to agree with more conservative political attitudes. More interestingly, Democrats residing in counties with higher exposure to the Show tend to be more moderate while Republicans in the same counties are less moderate. These results reveal that two potential mechanism on how Rush Limbaugh is impacting election outcomes through mobilisation. First, he not only by mobilises his own, strongly conservative audience to cast a vote to prevent the ideological ``enemy'' from winning. Second, Democrats living in counties with higher exposure to the Show tend to be more moderate and compared to stronger Democrats,  moderate Democrats have a lower probability to cast a vote on election day.

\begin{table}[htpb!]
\small
\caption{Exposure to The Show and Political Attitudes}
\label{tab:polar}
\begin{center}
\resizebox{1.09\textwidth}{!}{
\begin{tabular}{lccccccc} \toprule
& (1) & (2) & (3) & (4) & (5) & (6) \\
 & \textbf{Strong Cons.}  & \textbf{Strong Cons.} & \textbf{Cons./} & \textbf{Cons./} & \textbf{Moderate} & \textbf{Moderate}  \\
&  & & \textbf{ Strong Cons.}  & \textbf{Strong  Con.}   &  &  \\
 \midrule
 &  &  &     \\
$HHIacc_{c,s}$ & -0.2549 &  -1.7918*** &  -1.4724&  -2.4646***& 2.1152*** & 5.0478***\\
& (0.7099) & (0.5514) &(1.2392)&(0.9180) & (0.7610) &(1.0786) \\
$HHIacc_{c,s}$ $\times$ $Rep_{i}$ & & 2.8215*** &&2.3739** &  &-5.1831*** \\
& & (0.8819) && (1.1316)&  & (1.3426)\\

$HHIall_{c,s}$& 0.2253& 0.2335 &  -0.0179&-0.0011 & -0.3568** & 0.3682**\\
&(0.2183) & (0.1694) &(0.3817)&(0.1813) & (0.1820) & (0.1865) \\
&  &   &     \\
Observations & 302,914 & 302,914 & 302,914& 302,914&  302,914& 302,914 \\

\midrule
Geographic Controls & YES & YES & YES & YES & YES & YES\\
Demographic Controls & YES& YES & YES & YES & YES & YES\\
Individual Controls  & YES & YES & YES & YES & YES & YES\\
Year FE  & YES & YES & YES & YES & YES & YES \\
State FE  & YES & YES & YES & YES & YES & YES \\
$Rep_{i}$ & NO & YES & NO & YES & NO & YES \\
\bottomrule
\multicolumn{7}{p{1.3\textwidth}}{\footnotesize{\textit{Notes:} Dependent variables are binary indicators equalling to one if the respondent's political views were strong conservative (Columns (1) - (2)), conservative/strong conservative (Columns (3) - (4)) or moderate (Columns (5) - (6)). All estimates include geographic (county area, elevation, ruggedness and their respective squared terms) and demographic (log of total population, population shares for male, White, Black, Asian, Hispanic and above 50 year categories, log of median earnings, Gini coefficient, poverty rate, unemployment and farming area) controls, as well as individual controls in the form of the respondent's age, race, gender, educational status, marital status and family income. $Rep_{i}$ is a binary indicator equalling to 1 if the respondent voted for the Republican party at the previous presidential election, and zero otherwise. The sample is limited to respondents identifying as Republican/Democratic voters. Standard errors, clustered at the county level, are in parenthesis. *** p$<$0.01, ** p$<$0.05, * p$<$0.1}} \\
\end{tabular}}
\end{center}
\end{table}

\begin{figure}[htpb!]
\caption{Exposure to The Show and Political Attitudes}
\label{fig:attitude}
\begin{center}
	\includegraphics[width=16cm,height=18cm,keepaspectratio]{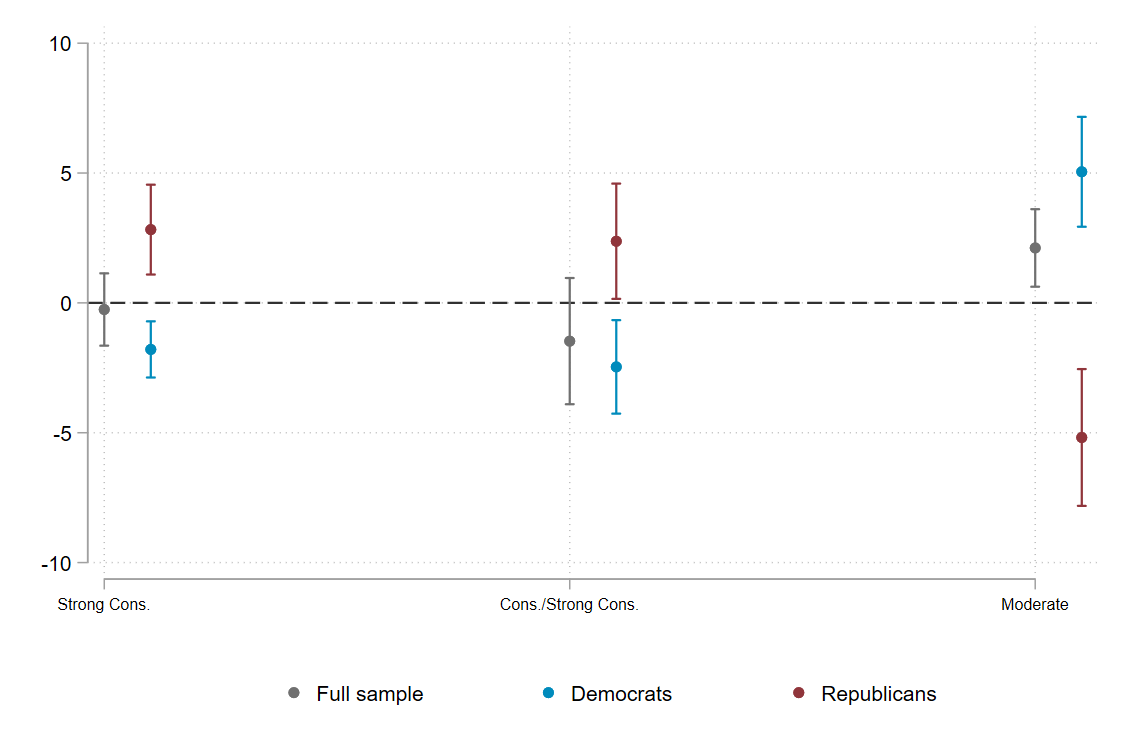}
	\end{center}
\captionsetup{width=\linewidth}
\caption*{\footnotesize{\textit{Notes:} Dots show the point estimates while the vertical lines depict the 95\% confidence intervals, based on standard errors clustered at the county level. Estimates for each political view/attitude category represent a separate regression estimate. All estimates include geographic (county area, elevation, ruggedness and their respective squared terms) and demographic (log of total population, population shares for male, White, Black, Asian, Hispanic and above 50 year categories, log of median earnings, Gini coefficient, poverty rate, unemployment and farming area) controls, as well as individual controls in the form of the respondent's age, race, gender, educational status, marital status and family income. The number of observations is 303,211.} }  
\end{figure}

\begin{figure}[htpb!]
\caption{Exposure to The Show and Political Attitudes - Heterogeneity across demographic groups}
\label{fig:attitude2}
\begin{center}
	\includegraphics[width=16cm,height=18cm,keepaspectratio]{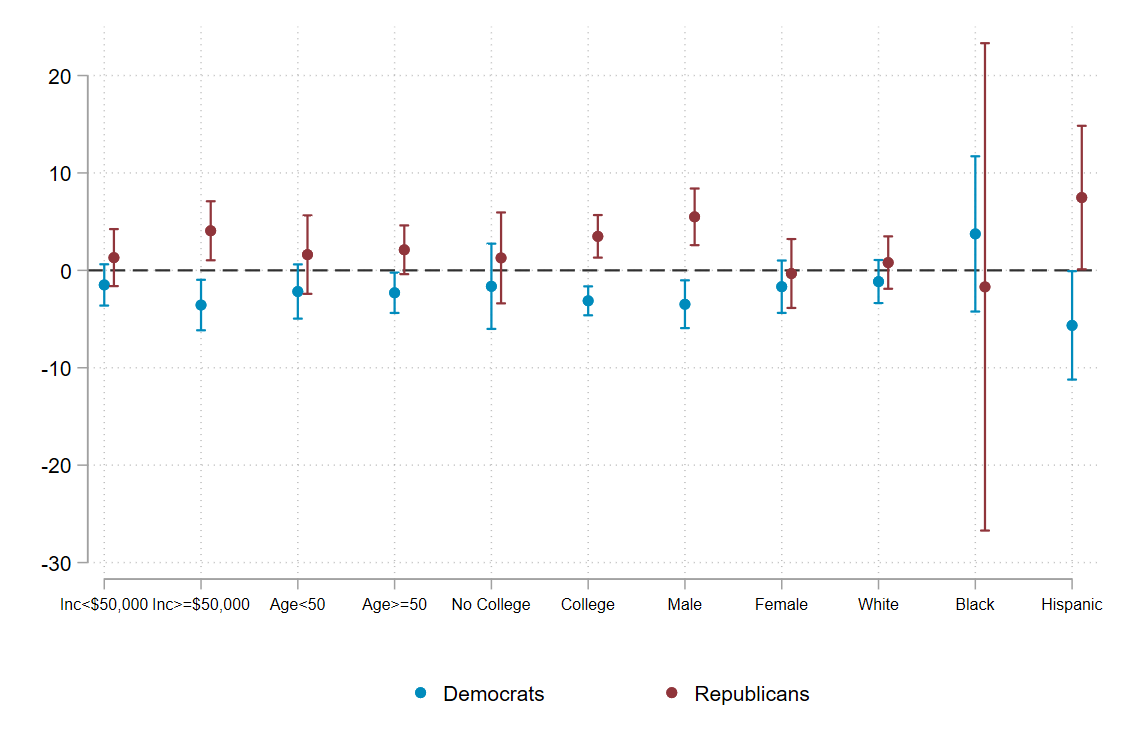}
	\end{center}
\captionsetup{width=\linewidth}
\caption*{\footnotesize{\textit{Notes:} Dots show the point estimates while the vertical lines depict the 95\% confidence intervals, based on standard errors clustered at the county level. Estimates for each demographic category represent a separate regression estimate. All estimates include geographic (county area, elevation, ruggedness and their respective squared terms) and demographic (log of total population, population shares for male, White, Black, Asian, Hispanic and above 50 year categories, log of median earnings, Gini coefficient, poverty rate, unemployment and farming area) controls, as well as individual controls in the form of the respondent's age, race, gender, educational status, marital status and family income.} } 
\end{figure}

In the next step, we analyse the heterogeneity of the effect by demographic groups. The results in Table \ref{tab:polar2} and Figure \ref{fig:attitude2} show the estimates for the binary outcome variable indicating if the respondent considers herself ``Conservative'' or ``Very Conservative''. Exposure to the Show appears to only have a statistically significant effect on respondents with a higher income level ($\geq$ \$50,000). Rush Limbaugh's audience was in general more politically knowledgeable and interested which is positively correlated with income. In general, the effect of exposure to the Show is also more pronounced for older and male respondents as well as people with college education. While we do not find systematic differences between white and African-American respondents, Hispanics who previously voted for Republicans seem to be the ethnic group most susceptible to the Show.

\begin{table}[htpb!]
\small
\caption{Heterogeneity}
\label{tab:polar2}
\begin{center}
\small
\resizebox{1.1\textwidth}{!}{
\begin{tabular}{lccccccccccc} \toprule
 & (1) & (2) & (3) & (4) & (5) & (6) & (7) & (8) & (9) & (10) & (11) \\
& Cons. & Cons. & Cons. & Cons. & Cons. & Cons. & Cons. & Cons. & Cons. & Cons. & Cons. \\
& Strong Cons. & Strong Cons. & Strong Cons. & Strong Cons. & Strong Cons. & Strong Cons. & Strong Cons. & Strong Cons. & Strong Cons. & Strong Cons. & Strong Cons. \\
\midrule
 &  &  &  &  &  &  &  &  &  &  &  \\
 & \multicolumn{2}{c}{Income} &  \multicolumn{2}{c}{Age} & \multicolumn{2}{c}{Education} & \multicolumn{2}{c}{Gender}  & \multicolumn{3}{c}{Race}\\
& $<$ \$50,000 & $\geq$ \$50,000 & $<$50 & $\geq$ 50 & No College & College & Male & Female & White & Black & Hispanic\\
  &  &  &  &  &  &  &  &  &  &  &  \\
$HHIacc_{c,s}$ & -1.4961 & -3.5625***& -2.1730  & -2.3044** &-1.6382 & -3.1298*** & -3.4778***& -1.6803 & -1.1525& 3.7397 &  -5.6498**\\
 & (1.0808) & (1.3210) & (1.4191) & (1.0550) & (2.2305)& (0.7577)  &(1.2508) & (1.3711) &(1.1276) & (4.0601)& (2.8365)  \\
$HHIacc_{c,s}$ $\times$ $Rep_{i}$ & 1.3064 & 4.0619*** & 1.6139 & 2.1102* &  1.2768  & 3.4901*** & 5.4924*** & -0.3264 & 0.7957 & -1.6967& 7.4703**\\
 & (1.4919) &(1.5447)  &  (2.0531)& (1.2750) & (2.3793) &(1.1132)  &(1.4828)  & (1.8028) & (1.3717) & (12.7484) & (3.7532) \\
$HHIall_{c,s}$ & 0.03671 & -0.0885 & -0.4669* & 0.4090* & 0.1818 & -0.1505 & -0.1204 &  0.0764&  -0.0522& -0.6349 & -0.9297 \\
 &(0.2366)  & (0.2668) & (0.2695) &  (0.2181)&  (0.4460)& (0.2075) & (0.2527) &(0.2504)   &  (0.2179)& (0.4656) &(0.8680)  \\
 
 &  &  &  &  &  &  &  &  &  &  &  \\
Observations & 109,826 & 161,003 & 105,034& 165,795 & 69,815 & 201,014 & 130,747 & 140,082 & 211,672 & 26,966 & 17,177 \\
\midrule
Geographic Controls & YES & YES & YES & YES & YES & YES & YES & YES & YES & YES & YES \\
Demographic Controls & YES & YES & YES & YES & YES & YES & YES & YES & YES & YES & YES \\
Individual Controls & YES & YES & YES & YES & YES & YES & YES & YES & YES & YES & YES \\
Year FE & YES & YES & YES & YES & YES & YES & YES & YES & YES & YES & YES \\
State FE & YES & YES & YES & YES & YES & YES & YES & YES & YES & YES & YES \\
Rep & YES & YES & YES & YES & YES & YES & YES & YES & YES & YES & YES \\
\bottomrule
\end{tabular}}
\end{center}
\end{table}

In the next step, we analysed the impact of the Show on individual preferences for particular policies.  Figure \ref{fig:attitude} presents the coefficients of point estimates of our exposure measure on five different policy questions for the full sample, Democrats and Republicans. We only find some indication that exposure to the show led to democrats adopting a strong Pro-Choice stance, and some impact on Republicans defending gun ownership.

\begin{figure}[htpb!]
\caption{Exposure to The Show and Policy Preferences  }
\label{fig:attitude}
\begin{center}
	\includegraphics[width=12cm,height=14cm,keepaspectratio]{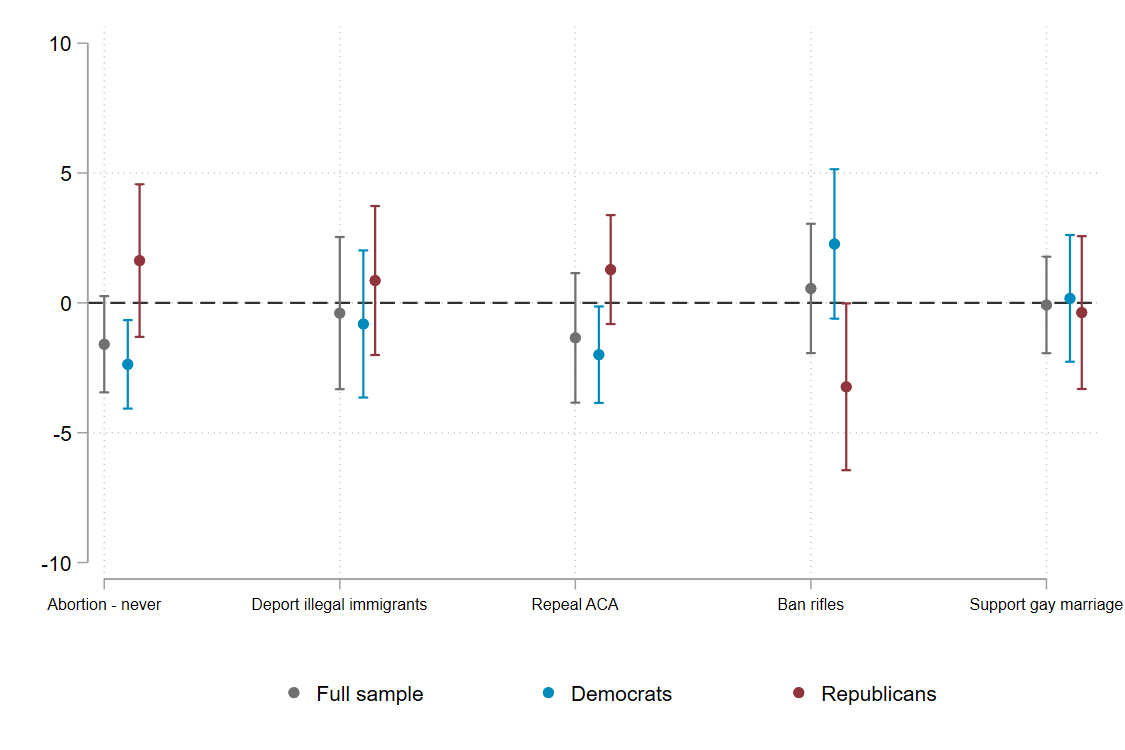}
	\end{center}
\captionsetup{width=\linewidth}
\caption*{\footnotesize{\textit{Notes:} Dots show the point estimates while the vertical lines depict the 95\% confidence intervals, based on standard errors clustered at the county level. EAll estimates include geographic (county area, elevation, ruggedness and their respective squared terms) and demographic (log of total population, population shares for male, White, Black, Asian, Hispanic and above 50 year categories, log of median earnings, Gini coefficient, poverty rate, unemployment and farming area) controls, as well as individual controls in the form of the respondent's age, race, gender, educational status, marital status and family income.} } 
\end{figure}

\subsection{ANES estimates based on Boxell et al.}\label{app: Online Appendix B}

Finally, we analyse the impact of exposure to the Rush Limbaugh Show on political polarisation following the methodolgy by \citet{boxell17}. We accessed the restricted version of the American National Election Studies (ANES) that allows us to link respondents place of residence with our county level exposure measures. Based on responses to the survey, we compute eight different measures of political polarization and an aggregate measure of political polarisation based on those eight individual measures.

The first two measures, ``Partisan affect polarization'' and ``Ideological affect polarization'', capture the respondents’ attitudes toward the members of the other political party. ``Partisan sorting'' and ``Partisan-Ideology polarization'' measures the difference between and individual's partisan identity and ideology. ``Perceived partisan-ideology polarization'' captures individual perception in ideological differences between Democrats and Republicans. ``Issue consistency'' and ``issue divergence'' capture how the respondent's issue positions line up on a single ideological dimension. Finally, ``Straight-ticket voting'' measures how often a respondent has split their votes across parties in an election. We then also calculate an ``Index'' that builds the average across those eight measures. 

The results are presented in Tables \ref{tab:polar1}, \ref{tab:polar2}, and \ref{tab:polar}.  In Table  \ref{tab:polar1} we show estimates that do not take into account the respondent's party affiliation. Similar to the results using the CCES data we only find some weak evidence that exposure to the Show increases political polarisation, mainly for Partisan-Ideology Polarisation and the overall Index. However, these results are only statistically significant at the 10\% level.

\begin{table}[htpb!]
\small
\caption{Effect of exposure to The Rush Limbaugh Show on measures of political polarization}
\label{tab:polar1}
\begin{center}
\small
\resizebox{1.1\textwidth}{!}{
\begin{tabular}{lccccccccc} \toprule
& (1) & (2) & (3) & (4) & (5) & (6) & (7) & (8) & (9) \\
 & \textbf{Partisan Affect} & \textbf{Ideological Affect} & \textbf{Partisan Sorting} & \textbf{Partisan-Ideology} & \textbf{Perceived Partisan-Ideology} & \textbf{Issue Consistency} & \textbf{Issue Divergence} & \textbf{Straight Ticket} & \textbf{Index}\\
 & \textbf{Polarization}  & \textbf{Polarization} & \textbf{Polarization} & \textbf{Polarization} & \textbf{Polarization} & & & \textbf{Voting}\\ \midrule
 &  &  &  &    \\
$HHIacc_{i}$ & -0.2914  & 1.5789&  0.1084 & 2.9010*&0.7865 & -1.0104& -1.6324**& 3.0375& 2.9124*\\
& (0.8851)  &(1.1359) & (1.5423)& (1.5350) & (1.2181)& (1.8345)& (0.8130)& (2.7175)& (1.6266)\\
$HHIall_{i}$ & 0.2653  &0.2575 & 0.5717&  0.4021&0.2652 & 0.6459& 0.0908& -0.8227& 0.8165 \\
& (0.1890)  & (0.3119) &(0.3676) & (0.5332) & (0.2550)& (0.4514)&(0.2305) & (0.9394)  & (0.5183)\\
&  &   &     \\
Observations & 16,809  & 9,486& 14,429&  9,285& 17,356&16,812 & 14,911&10,753 & 4,292\\
\midrule
Geographic Controls & YES & YES & YES & YES& YES & YES & YES & YES & YES\\
Demographic Controls & YES & YES & YES & YES& YES&  YES & YES & YES & YES\\
Individual Controls & YES & YES & YES & YES& YES & YES & YES & YES & YES\\
Year FE & YES & YES & YES & YES& YES & YES & YES & YES & YES\\
State FE & YES & YES & YES & YES& YES & YES & YES & YES & YES\\
\bottomrule
\end{tabular}}
\end{center}
\end{table}

Accounting for the potential differences in exposure between party affiliation, the results in Table \ref{tab:polar2}, show a clearer trend, revealing that exposure to the Show has a large impact on Republican voters. Republicans residing in counties with more exposure to the Show are more ideological aligned with their party and did more consistently vote for the Republican party in past elections.

\begin{table}[htpb!]
\small
\caption{Effect of exposure to The Rush Limbaugh Show on measures of political polarization - by party affiliation}
\label{tab:polar2}
\begin{center}
\small
\resizebox{1.1\textwidth}{!}{
\begin{tabular}{lccccccccc} \toprule
& (1) & (2) & (3) & (4) & (5) & (6) & (7) & (8) & (9) \\
 & \textbf{Partisan Affect} & \textbf{Ideological Affect} & \textbf{Partisan Sorting} & \textbf{Partisan-Ideology} & \textbf{Perceived Partisan-Ideology} & \textbf{Issue Consistency} & \textbf{Issue Divergence} & \textbf{Straight Ticket} & \textbf{Index}\\
 & \textbf{Polarization}  & \textbf{Polarization} & \textbf{Polarization} & \textbf{Polarization} & \textbf{Polarization} & & & \textbf{Voting}\\ \midrule
 &  &  &  &    \\
$HHIacc_{i}$ & -1.5549& -1.3436 & -4.7015***&8.1265*** & -0.6508 & -5.3733*&-3.6726***& 0.8629& -0.9538\\
& (1.7138) &(2.3459)  & (1.7923)& (2.4481)& (1.5651) & (2.7573)&(1.0386)&(4.1228) & (1.5627)\\
$HHIacc_{i}$ $\times$ $Rep$ &2.5593  & 4.4857* & 8.2601***& -7.8439***& 3.6864*** & 9.7452***&3.5403**& 7.6021**&2.1580 \\
& (2.0161) & (2.4019) & (2.4228)& (2.6565)& (1.4522) & (3.0924) &(1.4208)& (3.7366)&(2.0551)\\

$HHIall_{i}$ &0.2587  &  0.3046& 0.3855&0.4629 & 0.2399 & 0.5227&0.1041& -0.8309& 0.4540\\
& (0.1858) &(0.3201)  & (0.3873)& (0.4125)&(0.2730)  & (0.4750)&(0.2351)& (0.8880)&(0.2907) \\
&  &   &     \\
Observations & 16,809& 8,952 &13,115 &9,285& 15,625 &14,821 &14,821& 10,098  & 4,292\\

\midrule
Geographic Controls & YES & YES & YES & YES& YES & YES & YES & YES & YES\\
Demographic Controls & YES & YES & YES & YES& YES&  YES & YES & YES & YES\\
Individual Controls & YES & YES & YES & YES& YES & YES & YES & YES & YES\\
Year FE & YES & YES & YES & YES& YES & YES & YES & YES & YES\\
State FE & YES & YES & YES & YES& YES & YES & YES & YES & YES\\
\bottomrule
\end{tabular}}
\end{center}
\end{table}

\newpage
\section{Conclusion}
This paper examines the impact of radio talk show, particularly The Rush Limbaugh Show, on election outcomes and political polarization in the U.S. We introduce a new indicator that can capture county-level exposure to the Show, based on a spatial HHI of radio frequencies. 
While competition within a radio market itself could be endogeneous to the political preferences of a county, we build a measure of radio frequency competition based on accidental frequency overlaps in a county. The identifying assumption is that, conditional on the overall level of radio frequency competition in a county, the variation in radio frequency competition from accidental contour overlaps is not systematically correlated to variation in unobservables that affect election and political attitudes.

We then combine this competition measure with county level political outcomes, specifically the Republican vote share, and find that counties with more intense exposure to the Show have a systematically higher Republican vote share. This effects becomes economically and statistically significant after the year 2000, in line with the rise of more populist groups within the Republican party. Next we combine these indicators of exposure with individual level attitudes on societal issues extracted from the CES, based on the information on the geo-location of survey respondents. We find that individuals located in counties with higher exposure to The Show express stronger anti-abortion, anti-gay marriage, anti-immigration, anti-gun control and anti-environmental regulation attitudes. These effects are more prominent for those self-identifying as Republicans. These results highlight that conservative talk radio have had a non-trivial effect on the rise of populism in the U.S. in recent years. In the next steps of this study, we will be exploring the mechanisms driving these effects.

\pagebreak

\bibliographystyle{ecca}
\bibliography{bibliothek}

\bigskip

\clearpage

\setcounter{footnote}{0}

\newpage \appendix \numberwithin{equation}{section} 

\begin{center}

{\LARGE Online Appendix}

\bigskip
{\LARGE Competing for Attention -- The Effect of Talk Radio on US Political Polarization }

\bigskip

By Ashani Amarasinghe\footnote{SoDa Laboratories, Monash University. Email: ashani.amarasinghe@monash.edu.}  Paul A. Raschky\footnote{Department of Economics  and SoDa Laboratories, Monash University; email: paul.raschky@monash.edu.} 

\bigskip
\end{center}

\setcounter{subsection}{0}
\renewcommand{\thesubsection}{\Alph{subsection}}
\setcounter{figure}{0}
\renewcommand{\thefigure}{\thesubsection.\arabic{figure}}

\setcounter{table}{0}
\renewcommand{\thetable}{\thesubsection.\arabic{table}}

\subsection{Robustness Checks} \label{app: Online Appendix A}

\begin{figure}[htpb!]
\caption{Spatial clustering of standard errors}
\label{fig:overtime_conley}
\begin{center}
	\includegraphics[width=12cm,height=14cm,keepaspectratio]{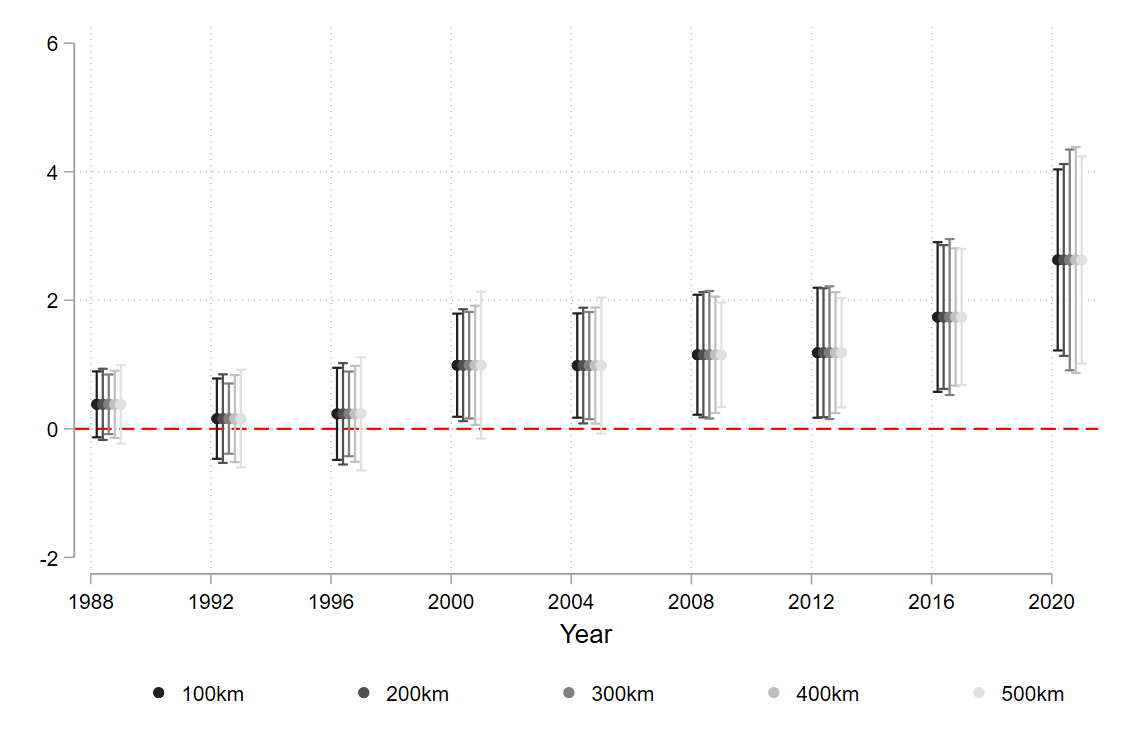}
	\end{center}
\captionsetup{width=\linewidth}
\caption*{\footnotesize{\textit{Notes:} Figure shows the effect of $HHIacc$ on Republican vote share over time. Dots show the point estimates while the vertical lines depict the 95\% confidence intervals, based on standard errors adjusted for spatial autocorrelation at 100, 200, 300, 400 and 500 km, respectively. Estimates for each election year and distance cutoff represent a separate regression estimate. All estimates include geographic (county area, elevation, ruggedness and their respective squared terms) and demographic (log of total population, population shares for male, black, Asian, Hispanic and above 50 year categories, log of median earnings, Gini coefficient, poverty rate, unemployment and farming area) controls. Estimates also control for the average Republican vote share over 1968-1984.} }  
\end{figure}

\begin{table}[htpb!]
\small
\caption{Effect of exposure to The Rush Limbaugh Show on pre-1988 Republican vote share}
\label{tab:election_placebo}
\begin{center}
\small
\resizebox{0.9\textwidth}{!}{
\begin{tabular}{lccccc} \toprule
 & (1) & (2) & (3) & (4) & (5) \\
& \multicolumn{5}{c}{\textbf{Dependent Variable:} \textit{1968-1984 Republican Vote Share$_{c,s}$}}\\ \midrule
 &  &  &  &  \\
$HHIacc_{c,s}$ &  1.9477*** & 0.4823 & 0.8808 &0.8209 & 0.7902 \\
 & (0.6392)& (0.4504) & (0.5522) & (0.5392)& (0.5338)  \\
$HHIall_{c,s}$ &  &  &  -0.2859&  -0.4334&  -0.1397\\
 &  &  & (0.2618)& (0.2719) & (0.2368)\\
 &  &  &  &    \\
Observations &  1,898 & 1,898 & 1,898 & 1,885 & 1,885 \\

\midrule

State FE & NO & YES & YES & YES & YES \\
Geographic Controls & NO & NO & NO & YES & YES \\
Demographic Controls & NO & NO & NO & NO & YES \\ \bottomrule
\multicolumn{6}{p{0.9\textwidth}}{\scriptsize{\textit{Notes:} The dependent variable is the average Republican vote share in the presidential elections over 1968-1984. Geographic controls include county area, elevation, ruggedness and their respective squared terms. Demographic controls include log of total population, population shares for male, black, Asian, Hispanic and above 50 year categories, log of median earnings, Gini coefficient, poverty rate, unemployment and farming area. Standard errors, clustered at the State level, are in parenthesis. *** p$<$0.01, ** p$<$0.05, * p$<$0.1}} \\
\end{tabular}}
\end{center}
\end{table}

\begin{table}[htpb!]
\small
\caption{Effect of exposure to The Rush Limbaugh Show on Republican vote share - Alternative specifications}
\label{tab:election_alt}
\begin{center}
\small
\resizebox{0.8\textwidth}{!}{
\begin{tabular}{lccc} \toprule
 & (1) & (2) & (3)  \\
 & Accidental = & Accidental =& Excluding \\
 & $<$25th percentile of & $<$average of & areas with low\\
 & overlap areas & overlap areas & nighttime light\\
 \midrule
 & \multicolumn{3}{c}{\textbf{A: Dependent Variable}}\\
& \multicolumn{3}{c}{\textit{2016 Republican Vote Share$_{c,s}$}}\\ \midrule
 &  &  &    \\
$HHIacc_{c,s}$ &   1.7914*** & 1.6887**  & 1.7507***   \\
 & (0.6614) & (0.6405)   & (0.6499)  \\
$HHIall_{c,s}$ &  -0.3630 &  -0.3509 &  -0.3581 \\
& (0.4086) & (0.4101)   & (0.4090)  \\
 &  &   &   \\
Observations & 1,885 &  1,885& 1,885\\

\midrule

 & \multicolumn{3}{c}{\textbf{B: Dependent Variable:}}\\
& \multicolumn{3}{c}{\textit{2020 Republican Vote Share$_{c,s}$}}\\ \midrule
 &  &  &   \\
$HHIacc_{c,s}$ &  2.6819***  &  2.6165*** &  2.6357***  \\
& (0.9502) & (0.9095)  &  (0.9346) \\
$HHIall_{c,s}$ &-0.7639  &  -0.7594 &   -0.7588\\
&(0.4772)  & (0.4746)  & (0.4768)  \\
 &  &  & \\
Observations & 1,796 & 1,796 & 1,796 \\
\midrule
State FE & YES & YES & YES \\
Geographic Controls & YES & YES & YES \\
Demographic Controls  & YES & YES & YES \\
Avg. Rep. Share 1968-1980 & YES & YES & YES \\ 
\bottomrule
\multicolumn{4}{p{0.87\textwidth}}{\scriptsize{\textit{Notes:} The dependent variable in Panels A and B is the Republican vote share in the 2016 and 2020 presidential elections, respectively. In Column (1) an accidental FM contour is defined as one where the overlap area between the FM contour and county is less than the 25th percentile of all such overlapping areas, while in Column (2) this cutoff is based on the average value of overlapping areas. In Column (3) we exclude accidental FM contours with low (less than the 10th percentile) nighttime light values. Geographic controls include county area, elevation, ruggedness and their respective squared terms. Demographic controls include log of total population, population shares for male, black, Asian, Hispanic and above 50 year categories, log of median earnings, Gini coefficient, poverty rate, unemployment and farming area. Standard errors, clustered at the State level, are in parenthesis. *** p$<$0.01, ** p$<$0.05, * p$<$0.1}} \\
\end{tabular}}
\end{center}
\end{table}

\begin{table}[htpb!]
\small
\caption{Effect of exposure to The Rush Limbaugh Show on Republican vote share - Spatial autoregressive and Spatial Durbin models}
\label{tab:election_spatiallags}
\begin{center}
\small
\resizebox{0.8\textwidth}{!}{
\begin{tabular}{lccc} \toprule
 & (1) & (2) \\
  & Spatial Autoregressive  Model & Spatial Durbin Model
\\
 \midrule
 & \multicolumn{2}{c}{\textbf{A: Dependent Variable}}\\
& \multicolumn{2}{c}{\textit{2016 Republican Vote Share$_{c,s}$}}\\ \midrule
 &  &  &    \\
$HHIacc_{c,s}$ &  1.7358** & 1.7639** \\
& (0.6593) & (0.6656) \\
$HHIall_{c,s}$ &  -0.3645 &   -0.3754 \\
& (0.4099) &(0.4051)  \\
$Neighb$ $Republican$ $Vote$ $Share_{c,s}$ & 0.0401* &0.0403*\\
 & (0.0204) &  (0.0200)   \\
 &  &   &   \\
Observations & 1,885 &  1,885 \\

\midrule

 & \multicolumn{2}{c}{\textbf{B: Dependent Variable}}\\
& \multicolumn{2}{c}{\textit{2020 Republican Vote Share$_{c,s}$}}\\ \midrule
 &  &  &    \\
$HHIacc_{c,s}$ & 2.6299*** &  2.5976*** \\
& (0.9333) & (0.9132) \\
$HHIall_{c,s}$ &  -0.7580 &  -0.7752\\
& (0.4768) & (0.4774) \\
$Neighb$ $Republican$ $Vote$ $Share_{c,s}$  & -0.0040&  -0.0065 \\
 & (0.0233) & (0.0246)    \\
 &  &   &   \\
Observations & 1,796 &  1,796\\

\midrule
State FE & YES &  YES\\
Geographic Controls & YES &  YES\\
Demographic Controls  & YES &  YES\\
Avg. Rep. Share 1968-1984 & YES &  YES\\
Spatial Lag of Dep. Var. & YES &  YES\\
Spatial Lag of Indep. Var. & NO & YES \\
Spatial Lag of Geographic Controls & NO & YES \\
Spatial Lag of Demographic Controls & NO & YES \\

\bottomrule
\multicolumn{3}{p{0.87\textwidth}}{\scriptsize{\textit{Notes:} The dependent variable in Panels A and B is the Republican vote share in the 2016 and 2020 presidential elections, respectively. Column (1) presents the estimates of the spatial autoregressive model, which includes the spatial lag of the dependent variable as a control variable. Column (2) presents the estimates of the spatial Durbin model which includes the spatial lag of the dependent variable as well as spatial lags of all independent variables, including controls. Spatial lags are based on the contiguity network of connectivity. Geographic controls include county area, elevation, ruggedness and their respective squared terms. Demographic controls include log of total population, population shares for male, black, Asian, Hispanic and above 50 year categories, log of median earnings, Gini coefficient, poverty rate, unemployment and farming area. Standard errors, clustered at the State level, are in parenthesis. *** p$<$0.01, ** p$<$0.05, * p$<$0.1}} \\
\end{tabular}}
\end{center}
\end{table}

 \end{document}